\documentclass[12pt,onecolumn,doublespace]{elsart_1}

\usepackage{graphicx}
\usepackage{graphics} 
\usepackage{epsfig} 
\usepackage{mathptmx} 
\usepackage{times} 
\usepackage{amsmath} 
\usepackage{amssymb}  
\usepackage{enumerate}
\usepackage{mathrsfs}
\usepackage{multirow}
\usepackage{subfigure}
\usepackage{latexsym}
\usepackage{bm}
\usepackage{dsfont}
\usepackage{multicol}
\usepackage{multirow}
\usepackage{color}
\usepackage{url}
\usepackage{algorithmic}
\usepackage{algorithm}
\usepackage{array}
\usepackage[algo2e]{algorithm2e}

{
	\begin{bmatrix}}%
	{\end{bmatrix}
	}
\newtheorem{proposition}{Proposition}

\newtheorem{theorem}{Theorem}

\newtheorem{assumption}{Assumption}

\graphicspath{{Figures/}}

\usepackage{color}

\newcommand{\MF}[1]{\textcolor{black}{#1}}

\makeatletter
\renewcommand\paragraph{\@startsection{paragraph}{4}{\z@}%
	{3.25ex \@plus1ex \@minus.2ex}%
	{-1em}%
	{\normalfont\normalsize}}
\makeatother
\begin{document}
	\begin{frontmatter}
		\title{A hierarchical multirate MPC scheme for interconnected systems\\ \emph{extended version}}
		\author[First]{M. Farina}
		\author[First]{X. Zhang}
		\author[First]{R. Scattolini}
		\address[First]{Dipartimento di Elettronica, Informazione e Bioingegneria, Politecnico di Milano, Milan 20133, Italy (e-mail:~\{marcello.farina,xinglong.zhang,riccardo.scattolini\}@polimi.it).}
		\begin{abstract}                
			This paper presents a hierarchical control scheme for interconnected linear systems. At the higher layer of the control structure a robust centralized Model Predictive Control (MPC) algorithm based on a reduced order dynamic model of the overall system optimizes a long-term performance index penalizing the deviation of the state and the control input from their nominal values. At the lower layer local MPC regulators, possibly working at different rates, are designed for the full order models of the subsystems to refine the control action computed at the higher layer. A simulation experiment is presented to describe the implementation aspects and the potentialities of the proposed approach.
		\end{abstract}
		
		\begin{keyword}
			Hierarchical MPC, multi-rate control, robust MPC, multivariable systems.
		\end{keyword}
		
	\end{frontmatter}
	%
	\section{Introduction and main idea} \label{Sec:Intro} 

\begin{figure*}[ht]
		\center
		\includegraphics[width=0.6\linewidth]{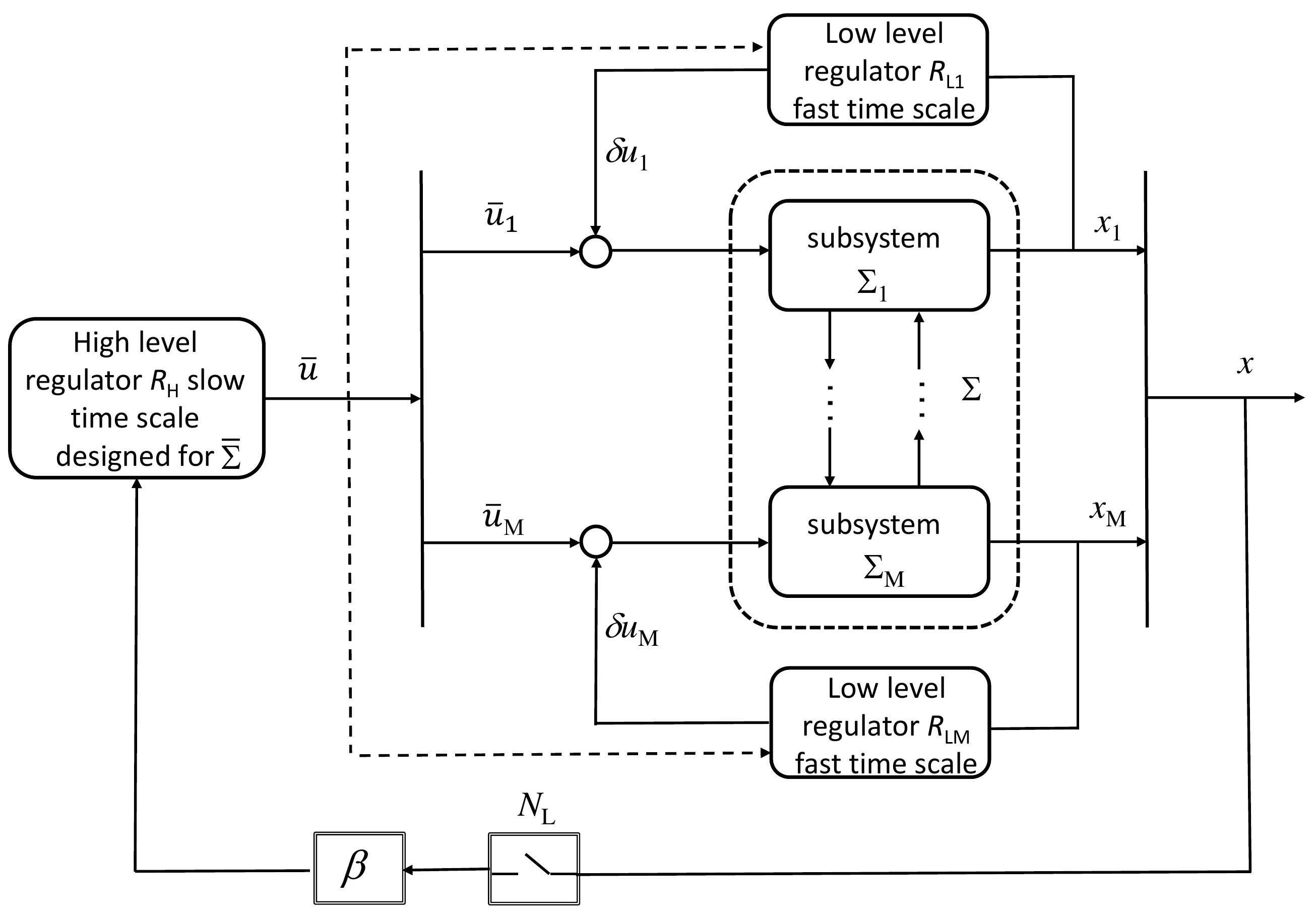}
		\caption{Overall control scheme.}
		\label{fig:GS}
	\end{figure*}

Physical and cyber-physical systems are becoming  more and more complex, large-scale, and heterogeneous due to the growing opportunities provided by information technology in terms of computing power, transmission of information, and networking capabilities. As a consequence, also the management and control of these systems represent a problem of increasing difficulty and require innovative solutions. A classical approach consists of resorting to decentralized or distributed control structures, see the fundamental book \cite{siljak2011} and, in the context of Model Predictive Control (MPC) here considered, the recent review \cite{Scattolini2009} on distributed MPC (DMPC) or the book \cite{negen}, where the most recent contributions to the design DMPC methods are reported. According to \cite{Scattolini2009}, DMPC algorithms can be cooperative, characterized by an intensive transmission load due to the multiple exchange of information among the regulators within one sampling period, or non-cooperative, characterized conservativeness to compensate for the effects of neglected dynamics. \\
With the aim of reducing the above limitations of DMPC algorithms, in this paper we propose the novel two-layer control scheme shown in Figure \ref{fig:GS}.
The system under control $\Sigma$ is assumed to be composed by $M$ interconnected subsystems $\Sigma_{1},...,\Sigma_{\rm \scriptscriptstyle M}$. A reduced order model $\bar{\Sigma}$ is first computed and a centralized MPC regulator $R_{\rm\scriptscriptstyle H}$ working at a slow rate is designed to consider the long-term behavior of the system and to compute the control variables $\bar{u}_i$, $i=1,...,M$. Then, faster local regulators $R_{{\rm\scriptscriptstyle L}i}$, $i=1,...,M$, are designed for each subsystem $\Sigma_{i}$ to compute the control contributions $\delta u_i$ compensating for the inaccuracies of the high layer design due to the mismatch between $\Sigma$ and $\bar{\Sigma}$. Notably, the local regulators can be designed and implemented at different rates to cope with subsystems operating at different time scales, as it often happens in many important industrial fields, see the centralized multirate MPC methods reported in \cite{Scatto88,Bequette1991,Lee1992,Christofides11,Christofides12a,Daoutidis02,Daoutidis12}, or the multirate implementations described in \cite{roshany2013kalman,heidarinejad2011multirate,zou2010network}.\\
A similar control structure has been already proposed in \cite{PicassoZhangScat} where however the framework was different: only independent systems $\Sigma_{i}$ with joint output constraints were considered, no multirate implementations were allowed, and the problem was to coordinate the $\Sigma_{i}'s$ to guarantee an overall output request, so that different technical tools, with respect to the ones here considered, had to be adopted at the two layers of the control structure.\\
	The paper is organized as follows. Section 2 introduces the models considered at the two layers. Section 3 describes the MPC algorithms adopted at the two layers, while Section 4 presents the main feasibility and convergence results as well as a summary of the main implementation aspects. Section 5 describes the simulation example, while in Section 6 some conclusions are drawn. The proofs of the main results are reported in the Appendix.\\
	\textbf{Notation:} for a given set of variables $z_{i}\in{\mathbb{R}}^{q_{i}}$,
	$i=1,2,\dots,M$, we define the vector whose vector-components are $z_{i}$ in the following compact form: $(z_{1}, z_{2}, \cdots, z_{\rm\scriptscriptstyle M})=[\,z_{1}^T\ z_{2}^T\ \cdots\ z_{\rm\scriptscriptstyle M}^T\,]^T\in{\mathbb{R}}^{q}$, where $q= \sum_{i=1}^{M}q_{i}$.
	We use $\mathcal{N}_+$ to denote the set of positive integer numbers.
	The symbols $\oplus$/$\ominus$ denote the Minkowski sum/difference. We denote by $\|\cdot\|$ the Euclidean norm.
	Finally, a ball with radius $\rho_{\varepsilon_{i}}$ and centered at $\bar{x}$ in the ${\mathbb{R}}^{dim}$
	space is defined as follows
	\[
	\mathcal{B}_{\rho_{\varepsilon_{i}}}(\bar{x}):=\{x\in{\mathbb{R}}^{dim}:||x-\bar{x}||\leq\rho_{\varepsilon_{i}}\}
	\]
	\section{Models for the two-layer control scheme}
	\label{Sec:linear subsystem}
	In this section we present the model of the overall system and the simplified one used for high-level control.
	\subsection{Large-scale system model}
	In line with \cite{LunzeBook92}, we assume that the overall system $\Sigma$ is
	composed by $M$ discrete-time, linear, interacting subsystems described by
	\begin{equation}
		\Sigma_{i}:\ \left\{ \begin{array}{lcl}
			x_{i}(h+1)&=&A_{\rm\scriptscriptstyle L}^{ii}x_{i}(h)+B_{\rm\scriptscriptstyle L}^{ii}u_{i}(h)+E_{\rm\scriptscriptstyle L}^{i}s_{i}(h)\\
			y_i(h)&=&C_{\rm\scriptscriptstyle L}^{ii}x_{i}(h)\\
			z_{i}(h)&=&C_{\rm\scriptscriptstyle L}^{zi}x_{i}(h),
		\end{array}\right.\label{Eqn:LL}
	\end{equation}
	$i=1,2,\dots,M$, where $x_{i}\subseteq{\mathbb{R}}^{n_{i}}$, $u_{i}\in {\mathbb{R}}^{m_{i}}$, and $y_{i}\in {\mathbb{R}}^{p_{i}}$
	are the state, input, and output vectors, while $h$ is the discrete-time index in the basic time scale according to which the models are defined and the low level regulators will be designed.
	The interconnections among the subsystems $\Sigma_{i}$ are represented by the coupling input and output vectors $s_{i}\in{\mathbb{R}}^{p_{si}}$
	and $z_{i}\in{\mathbb{R}}^{p_{zi}}$, respectively, where
	\begin{equation}
		s_{i}(h)=\sum_{j=1}^{M}L_{ij}z_{j}(h)\label{Eqn:si}
	\end{equation}
	with $L_{ii}=0$, $i=1,...,M$.\\
	From~\eqref{Eqn:LL} and~\eqref{Eqn:si}, the overall dynamical model $\Sigma$ is
	\begin{equation}
		\Sigma:\,\left\{
		\begin{array}{lcl}x(h+1)&=&A_{\rm\scriptscriptstyle L}x(h)+B_{\rm\scriptscriptstyle L}u(h)\\
			y(h)&=&C_{\rm\scriptscriptstyle L}x(h)
		\end{array}\right.
		\label{Eqn:C full model}
	\end{equation}
	where $x=(x_{1},\dots,x_{\rm\scriptscriptstyle M})\in\mathbb{R}^{n}$, $n=\sum_{i=1}^{\rm\scriptscriptstyle M}n_{i}$,
	$u=(u_{1},\dots,u_{\rm\scriptscriptstyle M})\in\mathbb{R}^{m}$, $m=\sum_{i=1}^{M}m_{i}$, and $y=(y_{1},\dots,y_{\rm\scriptscriptstyle M})\in\mathbb{R}^{p}$, $p=\sum_{i=1}^{M}p_{i}$. The diagonal blocks of $A_{\rm\scriptscriptstyle L}$
	are state transition matrices $A_{\rm\scriptscriptstyle L}^{ii}$, whereas
	the coupling terms among the $\Sigma_i's$ correspond to the non-diagonal
	blocks of $A_{\rm\scriptscriptstyle L}$, i.e., $A_{\rm\scriptscriptstyle L}^{ij}=E_{\rm\scriptscriptstyle L}^{i}L_{ij}C_{\rm\scriptscriptstyle L}^{zj}$,
	with $j\neq i$. The collective input and output matrices are $B_{\rm\scriptscriptstyle L}=$diag$(B_{\rm\scriptscriptstyle L}^{11},...,\,B_{\rm\scriptscriptstyle L}^{\rm\scriptscriptstyle MM})$ and $C_{\rm\scriptscriptstyle L}=$diag$(C_{\rm\scriptscriptstyle L}^{11},...,\,C_{\rm\scriptscriptstyle L}^{\rm\scriptscriptstyle MM})$, respectively.\\
	Concerning systems \eqref{Eqn:LL} and \eqref{Eqn:C full model}, the
	following standing assumption is introduced:
	\begin{assumption}\label{eq:Assump_LL}\hfill
		\begin{enumerate}
			\item \label{eq:Assump_LL_statefeedback} The state $x_{i}$ is measurable, for each $i=1,\dots,M$;
			\item \label{eq:Assump_LL_Schur} $A_{\rm\scriptscriptstyle L}$ is Schur stable;
			\item \label{eq:Assump_invariant_zero} $m=p$ and the system matrix
			$S_{\Sigma}=\begin{bmatrix}
			I-A_{\rm\scriptscriptstyle L}&-B_{\rm\scriptscriptstyle L}\\
			C_{\rm\scriptscriptstyle L}&0
			\end{bmatrix}$
			has full rank $n+m$;
			\item $B_{\rm\scriptscriptstyle L}$ and $C_{\rm\scriptscriptstyle L}$ are full-rank matrices;
			\item \label{eq:Assump_LL_Reach} the pair $(A_{\rm\scriptscriptstyle L}^{ii},\,B_{\rm\scriptscriptstyle L}^{ii})$ is
			reachable, for each $i=1,\dots,M$.\hfill$\square$
	\end{enumerate}\end{assumption}
	\subsection{Reduced order models}
	For each subsystem $\Sigma_{i},\,i=1,...,\,M$, we define a reduced order model $\bar{\Sigma}_{i},\,i=1,...,\,M$, with state
	$\bar{x}_{i}\in\mathbb{R}^{\bar{n}_{i}}$, $\bar{n}_i \leq n_i$, and input $\bar{u}_{i}\in \mathbb{R}^{m_i}$.
	%
	%
	%
	In a collective form, these systems $\bar{\Sigma}_i$
	define the overall reduced order model
	\begin{equation}
		\bar{\Sigma}:\,\left\{ \begin{array}{clc}
			\bar{x}(h+1)&=&A_{\rm\scriptscriptstyle H}\bar{x}(h)+B_{\rm\scriptscriptstyle H}\bar{u}(h)\\
			\bar{y}(h)&=&C_{\rm\scriptscriptstyle H}\bar{x}(h)\end{array}\right.\label{Eqn:C HL model}
	\end{equation}
	where $\bar{x}=(\bar{x}_{1},\dots,\bar{x}_{\rm\scriptscriptstyle M})\in\mathbb{R}^{\bar{n}}$,
	$\bar{n}=\sum_{i=1}^{M}\bar{n}_{i}$, $\bar{u}=(\bar{u}_{1},\dots,\bar{u}_{\rm\scriptscriptstyle M})\in\mathbb{R}^{m}$, and $\bar{y} \in \mathbb{R}^{p}$.
	The reduced order models $\bar{\Sigma}_i$ can be defined according to different criteria. First, it is necessary that the stability properties of system $\Sigma$ are inherited by $\bar{\Sigma}$. Moreover, we assume that for each subsystem $i=1,\dots,M$ there exists a state projection
	$\beta_i:\mathbb{R}^{{n}_{i}}\rightarrow \mathbb{R}^{\bar{n}_{i}}$, $i=1,...,M$, that allows to establish a connection between the states $x_i(h)$ of the original models and the states of the reduced models $\bar{x}_i(h)$. Collectively, we define $\beta=\text{diag}(\beta_1,...,\beta_{\rm\scriptscriptstyle M})$. In principle, the ideal case would be to verify that, if $\bar{u}(h)={u}(h)$, then $\bar{x}(h)=\beta x(h)$ and $\bar{y}(h)=y(h)$ for all $h\geq 0$ for suitable initial conditions. However, due to model reduction approximations, this ideal assumption must be relaxed; instead, we just ask that $\bar{x}=\beta x$ and that $\bar{y}=y$ in steady-state conditions. We also require that, while matrix $B_{\rm\scriptscriptstyle H}$ can be full, the output matrix $C_{\rm\scriptscriptstyle H}$ preserves the block-diagonal form of $C_{\rm\scriptscriptstyle L}$, i.e., that $C_{\rm\scriptscriptstyle H}=$diag$(C_{\rm\scriptscriptstyle H}^{11},\dots,C_{\rm\scriptscriptstyle H}^{\rm\scriptscriptstyle MM})$, where $C_{\rm\scriptscriptstyle H}^{ii}\in\mathbb{R}^{p_i\times \bar{n}_i}$ for all $i=1,\dots,M$.\\
	Overall, we require the following standing assumption to be satisfied to guarantee the compatibility of the models used at the two layers.
	\begin{assumption}\label{Assump:3}\hfill
		\begin{enumerate}
			\item \label{Assump:3_0} $A_{\rm\scriptscriptstyle H}$ is Schur stable;
			\item \label{Assump:3_1} $\beta_i$ is full rank and is such that $C_{\rm\scriptscriptstyle H}^{ii}\beta_i=C_{\rm\scriptscriptstyle L}^{ii}$, for each $i=1,\dots,M$;
			\item \label{Assump:3_2} letting $\hat{G}_{\rm\scriptscriptstyle L}(z)=\beta (zI-A_{\rm\scriptscriptstyle L})^{-1}B_{\rm\scriptscriptstyle L}$ and $G_{\rm\scriptscriptstyle H}(z)=(zI-A_{\rm\scriptscriptstyle H})^{-1}B_{\rm\scriptscriptstyle H}$, it holds that ${G}_{\rm\scriptscriptstyle H}(1)$ is full rank and $\hat{G}_{\rm\scriptscriptstyle L}(1)={G}_{\rm\scriptscriptstyle H}(1)$.\hfill$\square$
	\end{enumerate}\end{assumption}
	An algorithm to compute the projections $\beta_i$ and the matrices of $\bar{\Sigma}$ is discussed in Appendix \ref{app:beta}, along the lines of~\cite{PicassoZhangScat}.
	\section{Design of the hierarchical control structure}
	In this section the regulators at the two layers of the hierarchical control structure are designed for the solution to a tracking control problem, i.e., to drive the output $y(h)$ of the system $\Sigma$ to the reference value ${y}_{\rm\scriptscriptstyle S}$, while respecting suitable input constraints.\\
	Thanks to Assumption~\ref{eq:Assump_LL}.(\ref{eq:Assump_invariant_zero}), it is possible to compute the reference pair $({x}_{\rm\scriptscriptstyle S},{u}_{\rm\scriptscriptstyle S})$ corresponding to ${y}_{\rm\scriptscriptstyle S}$, i.e., such that ${x}_{\rm\scriptscriptstyle S}=A_{\rm\scriptscriptstyle L}{x}_{\rm\scriptscriptstyle S}+B_{\rm\scriptscriptstyle L}{u}_{\rm\scriptscriptstyle S}$ and $C_{\rm\scriptscriptstyle L}{x}_{\rm\scriptscriptstyle S}={y}_{\rm\scriptscriptstyle S}$. Correspondingly, we define $\bar{u}_{\rm\scriptscriptstyle S}={u}_{\rm\scriptscriptstyle S}$ as the steady-state input reference for the reduced-order system $\bar{\Sigma}$, and the corresponding reference steady-state value as $\bar{x}_{\rm\scriptscriptstyle S}=G_{\rm\scriptscriptstyle H}(1)\bar{u}_{\rm\scriptscriptstyle S}=\beta {x}_{\rm\scriptscriptstyle S}$ by Assumption~\ref{Assump:3}.(\ref{Assump:3_2}).\\
	At the same time, we aim to enforce input constraints of type $u_i(h)\in\bar{u}_{{\rm\scriptscriptstyle S},i}\oplus \mathcal{U}_{{\rm\scriptscriptstyle S},i}$ for all $i=1, \dots, M$, where $\bar{u}_{{\rm\scriptscriptstyle S},i}$ is the $i$-th vector component of $\bar{u}_{{\rm\scriptscriptstyle S}}$ and $\mathcal{U}_{{\rm\scriptscriptstyle S},i}$, $i=1,\dots,M$, are closed and convex sets containing the origin in their interiors. Note that, if the reference $\bar{u}_{{\rm\scriptscriptstyle S},i}$ changes, also the set $\mathcal{U}_{{\rm\scriptscriptstyle S},i}$ may vary to enforce absolute input limitations or saturations. At a collective level, the required constraints are $u(h)\in\bar{u}_{\rm\scriptscriptstyle S}\oplus\mathcal{U}_{\rm\scriptscriptstyle S}$, where $\mathcal{U}_{\rm\scriptscriptstyle S}=\prod_{i=1}^M
	\mathcal{U}_{{\rm\scriptscriptstyle S},i}$ is a closed and convex set containing the origin in its interior.
	\subsection{Design of the high level regulator}
	\label{Sec:HLregulator}
	The high level regulator, designed to work at a low frequency, is based on the reduced order model \eqref{Eqn:C HL model} sampled
	with period $N_{\rm\scriptscriptstyle L}$ under the assumption that, $\forall\,k\in{\mathbb{N}}$,
	the $\bar{u}_{i}'s$ are held constant over the interval $h\in[kN_{\rm\scriptscriptstyle L},(k+1)N_{\rm\scriptscriptstyle L}-1]$. \MF{Therefore, the sampling time of the high-level model is $N_{\rm\scriptscriptstyle L}$ times larger than the basic sampling time, used in the model \eqref{Eqn:LL}.}
	Denoting by $\bar{u}_{i}^{{[N_{\rm\scriptscriptstyle L}]}}(k)$ the constant
	values of $\bar{u}_i$ in the long sampling period $k$ and by $\bar{u}^{{[N_{\rm\scriptscriptstyle L}]}}(k)$ the overall
	input vector, the reduced order model in the slow timescale is
	\begin{equation}
		\bar{\Sigma}^{{[N_{\rm\scriptscriptstyle L}]}}:\ \begin{array}{l}
			\bar{x}^{{[N_{\rm\scriptscriptstyle L}]}}(k+1)=A_{{\rm {\rm\scriptscriptstyle H}}}^{N_{\rm\scriptscriptstyle L}}\bar{x}^{{[N_{\rm\scriptscriptstyle L}]}}(k)+B_{{\rm {\rm\scriptscriptstyle H}}}^{{[N_{\rm\scriptscriptstyle L}]}}\bar{u}^{{[N_{\rm\scriptscriptstyle L}]}}(k)\end{array}\label{Eqn:C HLS model}
	\end{equation}
	where $B_{{\rm {\rm\scriptscriptstyle H}}}^{{[N_{\rm\scriptscriptstyle L}]}}=\sum_{j=0}^{N_{\rm\scriptscriptstyle L}-1}A_{{\rm {\rm\scriptscriptstyle H}}}^{j}B_{{\rm {\rm\scriptscriptstyle H}}}$. To enforce the input constraints specified above, we will require that $\bar{u}^{{[N_{\rm\scriptscriptstyle L}]}}(k)\in\bar{u}_{\rm\scriptscriptstyle S}\oplus\bar{\mathcal{U}}_{\rm\scriptscriptstyle S}$, where $\bar{\mathcal{U}}_{\rm\scriptscriptstyle S}= \prod_{i=1}^M
	\bar{\mathcal{U}}_{{\rm\scriptscriptstyle S},i}$, $\bar{\mathcal{U}}_{{\rm\scriptscriptstyle S},i}\subset {\mathcal{U}}_{{\rm\scriptscriptstyle S},i}$ for each $i=1,\dots,M$, and where the properties of sets $\bar{\mathcal{U}}_{{\rm\scriptscriptstyle S},i}$ are specified later in the paper.\\
	In order to feedback a value of $\bar{x}^{[N_{\rm\scriptscriptstyle L}]}$ related to the real
	state $x$ of the system, the projected value $\beta_{i}x_{i}(kN_{\rm\scriptscriptstyle L})$ is used, so that the reset
	\begin{equation}
		\bar{x}_{i}^{[N_{\rm\scriptscriptstyle L}]}(k)=\beta_{i}x_{i}(kN_{\rm\scriptscriptstyle L})\label{Eqn:betai}
	\end{equation}
	
	must be applied for all $i=1,\dots,M$. In collective form \eqref{Eqn:betai} becomes
	\begin{equation}\bar{x}^{[N_{\rm\scriptscriptstyle L}]}(k)=\beta x(kN_{\rm\scriptscriptstyle L})\label{eq:reset_bar}\end{equation}
	The reset \eqref{Eqn:betai} at time $k$
	may force $\bar{x}^{[N_{\rm\scriptscriptstyle L}]}(k+1)$ to assume a value different from
	the one computed based on the dynamics \eqref{Eqn:C HLS model}
	and the applied input $\bar{u}^{[N_{\rm\scriptscriptstyle L}]}(k)$. This discrepancy, due to the model reduction error and to the actions of the low level controllers,
	is accounted for by including in \eqref{Eqn:C HLS model} an additive disturbance $\bar{w}(k)$, i.e.,
	\begin{equation}
		\bar{\Sigma}_{w}^{{[N_{\rm\scriptscriptstyle L}]}}:\ \begin{array}{l}
			\bar{x}^{{[N_{\rm\scriptscriptstyle L}]}}(k+1)=A_{{\rm {\rm\scriptscriptstyle H}}}^{N_{\rm\scriptscriptstyle L}}\bar{x}^{{[N_{\rm\scriptscriptstyle L}]}}(k)+B_{{\rm {\rm\scriptscriptstyle H}}}^{{[N_{\rm\scriptscriptstyle L}]}}\bar{u}^{{[N_{\rm\scriptscriptstyle L}]}}(k)+\bar{w}(k)\end{array}\label{Eqn:C HLS model1}
	\end{equation}
	The size of $\bar{w}(k)$ depends on the action of the low level regulators and its presence requires to resort to a robust MPC method, which is here designed assuming that $\bar{w}(k)\in \mathcal{W}$, where $\mathcal{W}$ is a compact set containing the origin. The characteristics of $\mathcal{W}$ will be defined in the following once the low level regulators have been specified (see Section 4).\\
	The robust MPC algorithm is based on the scheme proposed in \cite{Mayne2005219}. To this end, we first need to define the ``unperturbed" prediction model
	\begin{equation}
		\bar{\Sigma}_{w}^{{[N_{\rm\scriptscriptstyle L}]},o}:\ \begin{array}{l}
			\bar{x}^{{[N_{\rm\scriptscriptstyle L}]},o}(k+1)=A_{{\rm {\rm\scriptscriptstyle H}}}^{N_{\rm\scriptscriptstyle L}}\bar{x}^{{[N_{\rm\scriptscriptstyle L}]},o}(k)+B_{{\rm {\rm\scriptscriptstyle H}}}^{{[N_{\rm\scriptscriptstyle L}]}}\bar{u}^{{[N_{\rm\scriptscriptstyle L}]},o}(k)\end{array}\label{Eqn:C HLS model_unp}
	\end{equation}
	and the control gain matrix $\bar{K}_{\rm\scriptscriptstyle H}$ such that, at the same time
	\begin{itemize}
		\item $F_{\rm\scriptscriptstyle H}=A_{{\rm {\rm\scriptscriptstyle H}}}^{N_{\rm\scriptscriptstyle L}}+B_{{\rm {\rm\scriptscriptstyle H}}}^{{[N_{\rm\scriptscriptstyle L}]}}\bar{K}_{\rm\scriptscriptstyle H}$ is Schur stable.
		\item $F_{\rm\scriptscriptstyle L}^{[N_{\rm\scriptscriptstyle L}]}=A_{\rm\scriptscriptstyle L}^{N_{\rm\scriptscriptstyle L}}+B_{\rm\scriptscriptstyle L}^{[N_{\rm\scriptscriptstyle L}]}\bar{K}_{\rm\scriptscriptstyle H}\beta$ is Schur stable, where $B_{\rm\scriptscriptstyle L}^{[N_{\rm\scriptscriptstyle L}]}=\sum_{j=0}^{N_{\rm\scriptscriptstyle L}-1}A_{{\rm {\rm\scriptscriptstyle L}}}^{j}B_{\rm\scriptscriptstyle L}$.
	\end{itemize}
	We define $\bar{e}(k)=\bar{x}^{{[N_{\rm\scriptscriptstyle L}]}}(k)-\bar{x}^{{[N_{\rm\scriptscriptstyle L}]},o}(k)$ and we let $\mathcal{Z}$ be a robust positively invariant (RPI) set - minimal, if possible - for the autonomous but perturbed system
	\begin{equation}
		\bar{\Sigma}_{w}^{{[N_{\rm\scriptscriptstyle L}]},e}:\ \begin{array}{l}
			\bar{e}(k+1)=F_{\rm\scriptscriptstyle H}\bar{e}(k)+\bar{w}(k)\end{array}\label{Eqn:C HLS model_err}
	\end{equation}
	\MF{The prediction horizon for the high-level MPC consists of $N_{\rm\scriptscriptstyle H}$ slow time steps}. Denoting by $\overrightarrow{\bar{u}^{{[N_{\rm\scriptscriptstyle L}]},o}}{(t:t+N_{\rm\scriptscriptstyle H}-1)}$ the sequence $\bar{u}^{{[N_{\rm\scriptscriptstyle L}]},o}(t)$, $\dots$, $\bar{u}^{{[N_{\rm\scriptscriptstyle L}]},o}(t+N_{\rm\scriptscriptstyle H}-1)$, at each slow time-step $t$ the following optimization problem is solved:
	\begin{equation}
		\begin{array}{l}
			\min_{\bar{x}^{{[N_{\rm\scriptscriptstyle L}]},o}(t),\overrightarrow{\bar{u}^{{[N_{\rm\scriptscriptstyle L}]},o}}{(t:t+N_{\rm\scriptscriptstyle H}-1)}}{J_{{\rm\scriptscriptstyle H}}\big(\bar{x}^{{[N_{\rm\scriptscriptstyle L}]},o}(t),\overrightarrow{\bar{u}^{{[N_{\rm\scriptscriptstyle L}]},o}}{(t:t+N_{\rm\scriptscriptstyle H}-1)})}\\
			\mbox{subject to:}\\
			\bullet\,\,\mbox{the unperturbed model dynamics}~\eqref{Eqn:C HLS model_unp}\\
			\bullet\,\,\mbox{the initial constraint}~\beta x(tN_{\rm\scriptscriptstyle L})-\bar{x}^{{[N_{\rm\scriptscriptstyle L}]},o}(t)\in \mathcal{Z}\\
			\bullet\,\,\mbox{the terminal constraint}~\bar{x}^{{[N_{\rm\scriptscriptstyle L}]},o}(t+N_{\rm\scriptscriptstyle H})\in \bar{x}_{\rm\scriptscriptstyle S}\oplus\bar{\mathcal{X}}_{\rm\scriptscriptstyle F}\\
			\bullet\,\, \bar{u}^{{[N_{\rm\scriptscriptstyle L}]},o}(k)\in{\bar{u}_{\rm\scriptscriptstyle S}\oplus\bar{\mathcal{U}_{\rm\scriptscriptstyle S}}\ominus \bar{K}_{\rm\scriptscriptstyle H}\mathcal{Z}},\,k=t,\dots,t+N_{\rm\scriptscriptstyle H}-1,
		\end{array}\label{Eqn:HLoptimiz_1}
	\end{equation}
	where the input constrained set has been properly tightened in accordance with the used tube-based control approach, and where	
\begin{equation}
		\begin{array}{cll}
			J_{{\rm\scriptscriptstyle H}}&=\sum_{k=t}^{t+N_{\rm\scriptscriptstyle H}-1}\|\bar{x}^{{[N_{\rm\scriptscriptstyle L}]},o}(k)-\bar{x}_{\rm\scriptscriptstyle S}\|_{Q_{\rm\scriptscriptstyle H}}^{2}+\|\bar{u}^{{[N_{\rm\scriptscriptstyle L}]},o}(k)-\bar{u}_{\rm\scriptscriptstyle S}\|_{R_{\rm\scriptscriptstyle H}}^{2}\\
			&+ \|\bar{x}^{{[N_{\rm\scriptscriptstyle L}]},o}(t+N_{\rm\scriptscriptstyle H})-\bar{x}_{\rm\scriptscriptstyle S}\|_{P_{\rm\scriptscriptstyle H}}^{2}
		\end{array}\label{Eqn:HL_cost}
	\end{equation}

	The set $\bar{\mathcal{X}}_{\rm\scriptscriptstyle F}$ is a positively invariant terminal set for the unperturbed system~\eqref{Eqn:C HLS model_unp} controlled with the stabilizing control law $\bar{u}^{{[N_{\rm\scriptscriptstyle L}]},o}(k)=\bar{K}_{\rm\scriptscriptstyle H} \bar{x}^{{[N_{\rm\scriptscriptstyle L}]},o}(k)$, satisfying $\bar{K}_{\rm\scriptscriptstyle H}\bar{\mathcal{X}}_{\rm\scriptscriptstyle F} \subseteq {\bar{\mathcal{U}_{\rm\scriptscriptstyle S}}\ominus \bar{K}_{\rm\scriptscriptstyle H}\mathcal{Z}}$. In view of this, $\bar{x}_{\rm\scriptscriptstyle S}\oplus\bar{\mathcal{X}}_{\rm\scriptscriptstyle F}$ results positively-invariant for~\eqref{Eqn:C HLS model_unp} controlled with the stabilizing auxiliary control law  $\bar{u}^{{[N_{\rm\scriptscriptstyle L}]},o}(k)=\bar{u}_{\rm\scriptscriptstyle S}+\bar{K}_{\rm\scriptscriptstyle H} (\bar{x}^{{[N_{\rm\scriptscriptstyle L}]},o}(k)-\bar{x}_{\rm\scriptscriptstyle S})$.\\
	The positive definite and symmetric weighting matrices ${Q_{\rm\scriptscriptstyle H}}$, ${R_{\rm\scriptscriptstyle H}}$ are free design parameters, while  ${P_{\rm\scriptscriptstyle H}}$ is computed as the solution to the Lyapunov equation
	\begin{equation}
		F_{\rm\scriptscriptstyle H}^T P_{\rm\scriptscriptstyle H} F_{\rm\scriptscriptstyle H}-P_{\rm\scriptscriptstyle H}=-(Q_{\rm\scriptscriptstyle H}+\bar{K}_{\rm\scriptscriptstyle H}^TR_{\rm\scriptscriptstyle H} \bar{K}_{\rm\scriptscriptstyle H})
		\label{eqn:HL_Lyap}
	\end{equation}
	Letting $\bar{x}^{{[N_{\rm\scriptscriptstyle L}]},o}(t|t),\overrightarrow{\bar{u}^{{[N_{\rm\scriptscriptstyle L}]},o}}{(t:t+N_{\rm\scriptscriptstyle H}-1|t)}$ be the solution to the optimization problem~\eqref{Eqn:HLoptimiz_1}, the control action, applied to system $\bar{\Sigma}_{w}^{[N_{\rm\scriptscriptstyle L}]}$ at time $t$, is
	\begin{equation}
		\bar{u}^{{[N_{\rm\scriptscriptstyle L}]}}(t)=\bar{u}^{{[N_{\rm\scriptscriptstyle L}]},o}(t|t)+\bar{K}_{\rm\scriptscriptstyle H}(\beta x(tN_{\rm\scriptscriptstyle L})-\bar{x}^{{[N_{\rm\scriptscriptstyle L}]},o}(t|t))
		\label{eqn:HL_control_input}
	\end{equation}
	%
	%
	\subsection{Design of the low level regulators}
	Recall that (see again Figure~\ref{fig:GS}) the overall control action to be applied to the real system $\Sigma$ has components generated by both the high-level and the low-level controllers, i.e.,
	\begin{equation}
		u_i(h)=\bar{u}_{i}^{[N_{\rm\scriptscriptstyle L}]}(\lfloor h/N_{\rm\scriptscriptstyle L} \rfloor)+\delta u_i(h)\label{contrcompl}
	\end{equation}
	The low level regulators are in charge of computing the local control corrections $\delta u_{i}\in{\mathcal{U}}_{{\rm\scriptscriptstyle S},i}\ominus{\bar{\mathcal{U}}}_{{\rm\scriptscriptstyle S},i}$ with the specific goal of compensating for the effect of the model inaccuracies at the high level expressed by the term $\bar{w}(k)$ in \eqref{Eqn:C HLS model1}.
	To this end, first define the auxiliary system $\hat{\Sigma}_{i}$: for $h=kN_{\rm\scriptscriptstyle L},\dots,(k+1)N_{\rm\scriptscriptstyle L}-1$
	\begin{equation}
		\hat{\Sigma}_{i}:\ \left\{ \begin{array}{lcl}
			\hat{x}_{i}(h+1)&=&A_{\rm\scriptscriptstyle L}^{ii}\hat{x}_{i}(h)+B_{\rm\scriptscriptstyle L}^{ii}\bar{u}_{i}^{{[N_{\rm\scriptscriptstyle L}]}}(\lfloor h/N_{\rm\scriptscriptstyle L}\rfloor)+E_{\rm\scriptscriptstyle L}^{i}\hat{s}_{i}(h)\\
			\hat{s}_{i}(h)&=&\sum_{j=1}^{M}L_{ij}\hat{z}_{j}(h)\\
			\hat{z}_{i}(h)&=&C_{\rm\scriptscriptstyle L}^{zi}\hat{x}_{i}(h)\\
			\hat{x}_{i}(kN_{\rm\scriptscriptstyle L})&=&x_{i}(kN_{\rm\scriptscriptstyle L})
		\end{array}\right.\label{Eqn:LL sdyn}
	\end{equation}
	and note that $\hat{\Sigma}_{i}$ can be simulated in a centralized way in the time interval $[kN_{\rm\scriptscriptstyle L},\,(k+1)N_{\rm\scriptscriptstyle L})$
	once the high level controller has computed $\bar{u}_{i}^{{[N_{\rm\scriptscriptstyle L}]}}(k)$ at the beginning of the time interval.
	
	Also denote by $\varDelta\Sigma_{i}$ the model given by the difference between systems~\eqref{Eqn:LL} and~\eqref{Eqn:LL sdyn}, with \eqref{Eqn:si} and \eqref{contrcompl}.
	\begin{equation}
		\varDelta\Sigma_{i}:\ \left\{ \begin{array}{lcl}
			\delta x_{i}(h+1)&=&A_{\rm\scriptscriptstyle L}^{ii}\delta x_{i}(h)+B_{\rm\scriptscriptstyle L}^{ii}\delta u_{i}(h)+E_{\rm\scriptscriptstyle L}^{i}\delta s_{i}(h)\\
			\delta{s}_{i}(h)&=&\sum_{j=1}^{M}L_{ij}\delta{z}_{j}(h)\\
			\delta z_{i}(h)&=&C_{\rm\scriptscriptstyle L}^{zi}\delta x_{i}(h)\\
			\delta x_{i}(kN_{\rm\scriptscriptstyle L})&=&0
		\end{array}\right.\label{Eqn:LL ddyn}
	\end{equation}
	where $\delta x_{i}(h)=x_{i}(h)-\hat{x}_{i}(h)$ , $\delta z_{i}(h)=z_{i}(h)-\hat{z}_{i}(h)$
	and $\delta s_{i}(h)=s_{i}(h)-\hat{s}_{i}(h)$.
	
\begin{figure}[ht]
	\center
	\includegraphics[width=0.95\columnwidth]{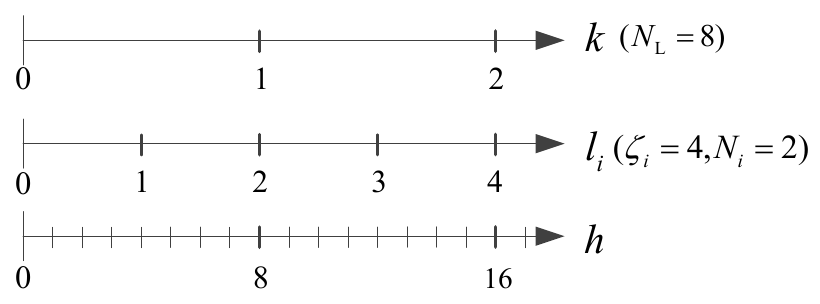}
		\caption{Adopted time scales: $k$ (high layer design), $l_i=kN_i$ (low layer design of the $i$-th local regulator), $h=\zeta_il_i=k\zeta_iN_i=kN_{\rm\scriptscriptstyle L}$ (basic time scale).}
		\label{fig:Timeaxis}
	\end{figure}

	The difference state $\delta x_{i}$ is available at each time instant
	$h$ since $x_{i}$ is measurable and $\hat{x}_{i}$ can be computed with \eqref{Eqn:LL sdyn}
	from the available value $\bar{u}_{i}^{{[N_{\rm\scriptscriptstyle L}]}}(\lfloor h/N_{\rm\scriptscriptstyle L}\rfloor)$.
	However, the difference dynamical system $\varDelta\Sigma_{i}$ is
	not yet useful for decentralized prediction since it depends upon the interconnection variables $\delta s_i(h)$ that, in turn, depend upon the variables $\delta x_j(h)$, $j\neq i$, not known in advance in the future prediction horizon. For this reason, we define a decentralized (approximated)
	dynamical system $\varDelta\hat{\Sigma}_{i}$ with input $\delta\hat{u}_i(h)$ and discarding all coupling inputs, i.e.,
	\begin{equation}
		\varDelta\hat{\Sigma}_{i}:\ \left\{
		\begin{array}{lcl}
			\delta\hat{x}_{i}(h+1)&=&A_{\rm\scriptscriptstyle L}^{ii}\delta\hat{x}_{i}(h)+B_{\rm\scriptscriptstyle L}^{ii}\delta \hat{u}_{i}(h)\\
			\delta\hat{x}_{i}(kN_{\rm\scriptscriptstyle L})&=&0
		\end{array}\right.\label{Eqn:LL dndyn}
	\end{equation}
	The decentralized dynamical system $\varDelta\hat{\Sigma}_{i}$ is suitable for prediction, since it does not depend on quantities related to other subsystems. However, the dynamics of the subsystems can be very different from each other and resampling can be advisable for the design of the low level regulators. To this end, for any subsystem $\Sigma_i$, define a new sampling period $\zeta_i\in \mathcal{N}_+$ such that $N_{\rm\scriptscriptstyle L}/\zeta_i=N_i\in \mathcal{N}_+$ and a corresponding time index $l_i$. For clarity, the relations among the time scales with indices $h$, $l_i$, and $k$ are shown in Figure \ref{fig:Timeaxis} in a specific case.\\

	In the time scale of $l_i$, define the dynamical system $\Delta\hat{\Sigma}_{i}^{[\zeta_i]}$ as
	\begin{equation}
		\varDelta\hat{\Sigma}_{i}^{[\zeta_i]}:\ \left\{ \begin{array}{c}
			\begin{array}{l}
				\delta\hat{x}_{i}^{[\zeta_i]}(l_i+1)=({A_{\rm\scriptscriptstyle L}^{ii}})^{\zeta_{i}}\delta\hat{x}_{i}^{[\zeta_i]}(l_i)+{B_{\rm\scriptscriptstyle L}^{ii}}^{[\zeta_i]}\delta \hat{u}^{[\zeta_i]}_{i}(l_i)\end{array}\\
			\delta\hat{x}_i^{[\zeta_i]}(kN_i)=\delta\hat{x}_i(kN_{\rm \scriptscriptstyle L})=0
		\end{array}\right.\label{Eqn:LL dndynopt}
	\end{equation}
	where ${B^{ii}_{\rm\scriptscriptstyle L}}^{[\zeta_{i}]}={\sum}_{j=0}^{\zeta_{i}-1}{(A^{ii}_{{\rm\scriptscriptstyle L}})}^{j}B^{ii}_{\rm\scriptscriptstyle L}$. In the short time scale, with time index $h$, the input $\delta\hat{u}_{i}(h)=\delta\hat{u}_{i}^{[\zeta_i]}(\lfloor h/\zeta_i \rfloor)$ is piecewise constant for
	$h \in [l_i\zeta_i,(l_{i}+1)\zeta_i-1]$. Its value will be computed as the result of a suitable optimization problem formulated for system \eqref{Eqn:LL dndynopt}.\\
	Given $\delta\hat{u}_{i}(h)$, the evolution of $\delta\hat{x}_{i}(h)$ can be computed thanks to the dynamical model  \eqref{Eqn:LL dndyn}, however  $\delta\hat{x}_{i}(h)$ is in general different from $\delta x_{i}(h)$ due to the neglected interconnections in systems~\eqref{Eqn:LL dndyn} and~\eqref{Eqn:LL dndynopt}. For this reason, and for all $i=1,\dots,M$, the input $\delta u_i(h)$ to the real model \eqref{Eqn:LL ddyn} is computed based on $\delta\hat{u}_{i}(h)$, $\delta x_i(h)$, and $\delta \hat{x}_i(h)$ using a standard state-feedback policy, i.e.,
	\begin{equation}
		\delta u_i(h)=\delta\hat{u}_{i}(h)+K_i (\delta {x}_i(h)-\delta \hat{x}_i(h))\label{Eqn:LL dndyn_sf_deltau}
	\end{equation}
	where $K_i$ is designed in such a way that the matrix $F_{\rm\scriptscriptstyle L}=A_{\rm\scriptscriptstyle L}+B_{\rm\scriptscriptstyle L}K$ is Schur stable, being $K=$diag$(K_1,\dots,K_{\rm\scriptscriptstyle M})$.\\
	Assume now to be at time $h=kN_{\rm\scriptscriptstyle L}$ and to have run the high level controller,
	so that both $\bar{u}_{i}^{{[N_{\rm\scriptscriptstyle L}]}}(k)$ and the
	predicted value $\bar{x}^{[N_{\rm\scriptscriptstyle L}]}(k+1|k)=A_{\rm\scriptscriptstyle H}^{N_{\rm\scriptscriptstyle L}}\bar{x}^{[N_{\rm\scriptscriptstyle L}]}(k)+B_{\rm\scriptscriptstyle H}^{[N_{\rm\scriptscriptstyle L}]}\bar{u}^{[N_{\rm\scriptscriptstyle L}]}(k)$ are available. Therefore,
	in order to remove the effect of the mismatch at the high level represented by $\bar{w}(k)$ in \eqref{Eqn:C HLS model1},
	the low level controller working in the interval $h\in[kN_{\rm\scriptscriptstyle L},(k+1)N_{\rm\scriptscriptstyle L}-1]$ should, if possible, aim to fulfill
	$$\beta_{i}x_{i}((k+1)N_{\rm\scriptscriptstyle L})=\bar{x}_{i}^{[N_{\rm\scriptscriptstyle L}]}(k+1|k)$$
	or equivalently,
	\begin{equation}
		\beta_{i}\delta x_{i}((k+1)N_{\rm\scriptscriptstyle L})=\bar{x}_{i}^{[N_{\rm\scriptscriptstyle L}]}(k+1|k)-\beta_{i}\hat{x}_{i}((k+1)N_{\rm\scriptscriptstyle L})\label{eq:terminal}
	\end{equation}
	
	Since the model used for low-level control design is the decentralized one (i.e., \eqref{Eqn:LL dndyn}), the constraint \eqref{eq:terminal} can only be formulated in an approximated way with reference to the state $\delta\hat{x}_{i}$ of system \eqref{Eqn:LL dndyn}. In turn, if the resampling is used with $\zeta_i\neq 1$, the constraint on $\delta\hat{x}_{i}$ must be reformulated in terms of the state $\delta\hat{x}_{i}^{[\zeta_i]}$ of system \eqref{Eqn:LL dndynopt}, so that
	\begin{equation}
		\begin{array}{rl}
		&\beta_{i}\delta\hat{x}_{i}^{[\zeta_i]}((k+1)N_i)=\beta_{i}\delta\hat{x}_{i}((k+1)N_{\rm\scriptscriptstyle L})=\\
		[0.1 cm]&=\bar{x}_{i}^{[N_{\rm\scriptscriptstyle L}]}(k+1|k)-\beta_{i}\hat{x}_{i}((k+1)N_{\rm\scriptscriptstyle L})\label{eq:terminaln}
		\end{array}
	\end{equation}
	Note that the fulfillment of \eqref{eq:terminaln} does not
	imply that \eqref{eq:terminal} is satisfied due to the neglected
	interconnections in \eqref{Eqn:LL dndyn} and \eqref{Eqn:LL dndynopt} which make the term $\bar{w}(k)$ in \eqref{Eqn:C HLS model} not identically equal to zero, although it contributes to its reduction.\\
	All these considerations lead to the formulation of the following low-level MPC designs. Letting $\overrightarrow{\delta \hat{u}^{[\zeta_i]}_i} (kN_i:(k+1)N_i-1)=(\delta \hat{u}^{[\zeta_i]}_{i}(kN_i), \dots, \delta \hat{u}^{[\zeta_i]}_{i}((k+1)N_i-1))\in({\mathbb{R}}^{m_{i}})^{{N_i}}$,
	the low level control action is computed, at time instant $h=kN_{\rm\scriptscriptstyle L}$, based on the solution to the following optimization problem:
	%
	\begin{equation}
		\begin{array}{l}
			\min_{\overrightarrow{\delta \hat{u}^{[\zeta_i]}_i} {(kN_i:(k+1)N_i-1)}} J_{{\rm\scriptscriptstyle L}}\left(\overrightarrow{\delta \hat{u}^{[\zeta_i]}_i} {(kN_i:(k+1)N_i-1)}\right)\\
			\mbox{subject to:}\\
			\bullet\,\,\mbox{the dynamics}~\eqref{Eqn:LL dndynopt}\\
			\bullet\,\,\mbox{the terminal constraint}~\eqref{eq:terminaln}\\
			\bullet\,\,\delta \hat{u}^{[\zeta_i]}_{i}(kN_i+l_i)\in{\Delta\hat{\mathcal{U}}_{i}},\,l_i=0,\dots,N_i-1,
		\end{array}\label{Eqn:LLoptimiz_1}
	\end{equation}
	where $J_{{\rm\scriptscriptstyle L}}$ is a positive definite function with arguments $\delta \hat{x}^{[\zeta_i]}_i(kN_i+l_i)$, $\delta \hat{u}^{[\zeta_i]}_{i}(kN_i+l_i)$, $l_i=0, \dots, N_i-1$, e.g.,
	\begin{equation}
		J_{{\rm\scriptscriptstyle L}}=\sum_{l_i=0}^{N_i-1}\|\delta\hat{x}^{[\zeta_i]}_{i}(kN_i+l_i)\|_{Q_{i}}^{2}+\|\delta \hat{u}^{[\zeta_i]}_{i}(kN_i+l_i)\|_{R_{i}}^{2}
		\label{Eqn:LL_cost}
	\end{equation}
	and where a discussion on how to select the set ${\Delta\hat{\mathcal{U}}_{i}}$ is deferred to Appendix~\ref{app:rhos}.\\
	Finally, for $j=0,\dots,N_{\rm\scriptscriptstyle L}-1$, the control component  at each (fast) time instant $\delta u_i(kN_{\rm\scriptscriptstyle L}+j)$ is given by
	\begin{equation}\label{llcontr_1}
		\begin{array}{rl}
		&\delta u_i(kN_{\rm\scriptscriptstyle L}+j)=\delta\hat{u}_i(kN_{\rm \scriptscriptstyle L}+j|kN_{\rm\scriptscriptstyle L})\\
		&+K_i(\delta{x}_i(kN_{\rm\scriptscriptstyle L}+j)-\delta \hat{x}_i(kN_{\rm\scriptscriptstyle L}+j|kN_{\rm\scriptscriptstyle L}))
		\end{array}
	\end{equation}
	where
	\begin{equation}\label{llcontr_2}
		\delta\hat{u}_i(kN_{\rm \scriptscriptstyle L}+j|kN_{\rm\scriptscriptstyle L})=\delta\hat{u}^{[\zeta_i]}_i(kN_i+\lfloor j/\zeta_i\rfloor|kN_{\rm\scriptscriptstyle L})
	\end{equation}
\MF{A further clarification is finally due. At the low level, the controller is designed mostly to compensate for the effects of model inaccuracies during each long sampling time (i.e., that of the high-level controller). Therefore, the prediction horizon at low level coincides with one large sampling time of the slow high-level controller, i.e., corresponding to $N_{\rm\scriptscriptstyle L}$ fast sampling times. Due to resampling, the optimization problem computed at the beginning of each slow sampling time, i.e., at time $h=kN_{\rm\scriptscriptstyle L}$, has a prediction horizon of $N_i$ steps of lenght $\zeta_i$, where indeed $N_i\zeta_i=N_{\rm\scriptscriptstyle L}$. As a result of this optimization problem, the input sequence $\delta\hat{u}^{[\zeta_i]}_i(kN_i|kN_{\rm\scriptscriptstyle L})$, $\dots$, $\delta\hat{u}^{[\zeta_i]}_i((k+1)N_i-1|kN_{\rm\scriptscriptstyle L})$ is obtained, and at each fast sampling time $kN_{\rm\scriptscriptstyle L}+j$, $j=0,\dots,N_{\rm\scriptscriptstyle L}-1$, the real (low-level) input contribution \eqref{llcontr_1} is used in \eqref{contrcompl}. Note that, in this way, $\delta u_i(h)$ varies at each sampling time.\\
	In summary, the on-line implementation of the hierarchical control scheme here proposed consists of the following steps:
	\begin{itemize}
		\item at any long sampling time $k$ solve the centralized simplified slow time-scale MPC problem \eqref{Eqn:HLoptimiz_1} with cost \eqref{Eqn:HL_cost} and obtain the value of $\bar{u}^{[N_{\rm\scriptscriptstyle L}]}(k)$ using \eqref{eqn:HL_control_input};
		\item for any system $\Sigma_i$, at the beginning of any long sampling time $k$ solve the optimization problem \eqref{Eqn:LLoptimiz_1} with cost \eqref{Eqn:LL_cost} and compute the full sequence $\delta\hat{u}^{[\zeta_i]}_i(kN_i|kN_{\rm\scriptscriptstyle L})$, $\dots$, $\delta\hat{u}^{[\zeta_i]}_i((k+1)N_i-1|kN_{\rm\scriptscriptstyle L})$;
		\item for any system $\Sigma_i$, at any fast sampling time $kN_{\rm\scriptscriptstyle L}+j$ $j=0,\dots, N_{\rm\scriptscriptstyle L}-1$, compute the low-level contribution $\delta u_i(kN_{\rm\scriptscriptstyle L}+j)$ using \eqref{llcontr_1} and \eqref{llcontr_2};
    \item for any system $\Sigma_i$, at any fast sampling time $kN_{\rm\scriptscriptstyle L}+j$ the applied control value is obtained as the sum of the low-level and of the high level contributions, as in \eqref{contrcompl}.
	\end{itemize}}
	%
	%
	%
	%
	%
	\section{Properties and algorithm implementation}
	\label{sec:rec_feas}
	The main assumptions, the recursive feasibility and convergence properties of the optimization problems stated at the high and low levels are established in this section.
\subsection{Main assumptions and remarks}
\label{subsec:mainass}
Define
	\begin{equation}\kappa(N_{\rm\scriptscriptstyle L})=\|\mathcal{B}(N_{\rm\scriptscriptstyle L})\|\label{Eqn:Gkappa1}\end{equation}
	where
	$$\mathcal{B}(N_{\rm\scriptscriptstyle L})=\sum_{j=1}^{N_{\rm\scriptscriptstyle L}}A_{\rm\scriptscriptstyle H}^{N_{\rm\scriptscriptstyle L}-j}B_{\rm\scriptscriptstyle H}-\beta\sum_{j=1}^{N_{\rm\scriptscriptstyle L}}A_{\rm\scriptscriptstyle L}^{N_{\rm\scriptscriptstyle L}-j}B_{\rm\scriptscriptstyle L}$$
	Also, let $I_{si}=\begin{bmatrix}0_{\bar{n}_{i}\times \bar{n}_{1}}&\cdots& I_{\bar{n}_{i}}&\cdots&0_{\bar{n}_{i}\times\bar{n}_{M}}\end{bmatrix}$, ${\mathcal{A}}(N_{\rm\scriptscriptstyle L})=A_{{\rm {\rm\scriptscriptstyle H}}}^{N_{\rm\scriptscriptstyle L}}\beta-\beta A_{{\rm L}}^{N_{\rm\scriptscriptstyle L}}\in{\mathbb{R}}^{\bar{n}\times n}$,
	${\mathcal{R}(N_{\rm\scriptscriptstyle L})}=\big[\,B_{{\rm {\rm\scriptscriptstyle L}}}\ A_{{\rm {\rm\scriptscriptstyle L}}}B_{{\rm {\rm\scriptscriptstyle L}}}\ \cdots\ (A_{{\rm {\rm\scriptscriptstyle L}}})^{N_{\rm\scriptscriptstyle L}-1}B_{{\rm {\rm\scriptscriptstyle L}}}\,\big]$,
	$\rho_{{u}}$ be such that ${\mathcal{U}_{\rm\scriptscriptstyle S}}\subseteq\mathcal{B}_{\rho_{{u}}}(0)$.
	%
	%
We now introduce the following assumption.
	\begin{assumption}\label{Assump:term_constr}\hfill
		\begin{enumerate}
			\item\label{Assump:term_constr_1} $\|A_{\rm\scriptscriptstyle L}^{N_{\rm\scriptscriptstyle L}}\|<1$;
			\item\label{Assump:term_constr_rk}  letting ${\mathcal{R}_{i}}(N_i)=\begin{bmatrix}{B_{{\rm {\rm\scriptscriptstyle L}}}^{ii}}^{[\zeta_i]}&(A_{{\rm {\rm\scriptscriptstyle L}}}^{ii})^{\zeta_i}{B_{{\rm {\rm\scriptscriptstyle L}}}^{ii}}^{[\zeta_i]}&\cdots&(A_{{\rm {\rm\scriptscriptstyle L}}}^{ii})^{\zeta_i(N_{i}-1)}{B_{{\rm {\rm\scriptscriptstyle L}}}^{ii}}^{[\zeta_i]}\end{bmatrix}$
			 for each $i=1,...,M$, be the reachability matrix in $N_i$ steps associated to $((A_{{\rm {\rm\scriptscriptstyle L}}}^{ii})^{\zeta_i},{B_{{\rm {\rm\scriptscriptstyle L}}}^{ii}}^{[\zeta_i]})$,
			matrix
			\[
			{\mathcal{H}}_{i}(N_i)=\beta_{i}{\mathcal{R}}_{i}(N_i)\in{\mathbb{R}}^{\bar{n}_{i}\times N_im_{i}}
			\]
			is full-rank with minimum singular value $\underline{\sigma}_{\mathcal{H}_{i}(N_i)}>0$;
			\item\label{Assump:term_constr_kappa} letting $\rho_{\bar{u}}$ and $\rho_{\delta\hat{u}_{i}}$ be such that
			${\mathcal{\bar{U}}}\subseteq\mathcal{B}_{\rho_{\bar{u}}}(0)$
			and ${\Delta\hat{\mathcal{U}}_{i}}\supseteq\mathcal{B}_{\rho_{\delta\hat{u}_{i}}}(0)$,
			respectively, for any $i=1,...,M$ it holds that
			\begin{equation}\label{eq:rho_delta_u}
				\rho_{\delta\hat{u}_{i}}>\frac{\kappa(N_{\rm\scriptscriptstyle L})\rho_{\bar{u}}}{\sqrt{N_i}\underline{\sigma}_{\mathcal{H}_{i}(N_i)}}
			\end{equation}
			\item\label{Assump:term_constr_4} for each $i=1,\dots,M$
			\begin{equation}
				\chi_{i}(kN_{\rm\scriptscriptstyle L})=\frac{\sqrt{N_{\rm\scriptscriptstyle L}}\varrho_{u}\|{\mathcal{R}(N_{\rm\scriptscriptstyle L})}\|\|{\mathcal{A}}(N_{\rm\scriptscriptstyle L})\|}{(1-\|A_{\rm\scriptscriptstyle L}^{N_{\rm\scriptscriptstyle L}}\|)(\sqrt{N_i}\underline{\sigma}_{\mathcal{H}_{i}(N_i)}\rho_{\delta \hat{u}_{i}}-\kappa(N_{\rm\scriptscriptstyle L})\rho_{\bar{u}})}\leq1\label{Eqn:chidef}
			\end{equation}
			\item\label{Assump:term_constr_5} Define $\Delta\bar{\mathcal{U}}_i=\Delta\mathcal{U}_i(N_{\rm\scriptscriptstyle L}-1)$, and $\Delta\mathcal{U}_i(j)=\Delta\hat{\mathcal{U}}_i\oplus \mathcal{B}_{\rho_{\Delta{u}_i}(j)}(0)$ where $\rho_{\Delta{u}_i}(j)=\sum_{r=2}^j\|K_i I_{si} F_{\rm\scriptscriptstyle L}^{j-r}(A_{\rm\scriptscriptstyle L}-A_{\rm\scriptscriptstyle L}^D)\|\rho_{\delta\hat{x}}(r-1)$ for all $j=2,\dots,N_{\rm\scriptscriptstyle L}$, $\rho_{\Delta{u}_i}(j)=0$ for $j=0,1$, and where $A_{\rm\scriptscriptstyle L}^{\rm\scriptscriptstyle D}=$diag$(A_{\rm\scriptscriptstyle L}^{11},\dots,A_{\rm\scriptscriptstyle L}^{\rm\scriptscriptstyle MM})$ and $\rho_{\delta\hat{x}}(j)= \sqrt{\sum_{i=1}^M\rho^2_{\delta\hat{x}_i}(j)}$,
			\begin{equation}\rho_{\delta\hat{x}_i}(j)=\rho_{\delta \hat{u}_i}\sum_{r=1}^{j}\|(A_{\rm\scriptscriptstyle L}^{ii})^{j-r}B_{\rm\scriptscriptstyle L}^{ii}\|\label{Eqn:bound_deltaxhat}\end{equation}
			We require that
			\begin{equation}\bar{\mathcal{U}}_{\rm\scriptscriptstyle S}\oplus(\prod_{i=1}^M \Delta \bar{\mathcal{U}}_i)\subseteq \mathcal{U}_{\rm\scriptscriptstyle S}\label{eq:bound u_def}\end{equation}\label{Assumption_bound_u_def} \hfill$\square$
		\end{enumerate}
	\end{assumption}
Assumption \ref{Assump:term_constr} may be viewed, at a first glance, as a list of purely abstract and technical requirements. However, on the one hand, it represents relevant inherent properties of the system under control and, on the other hand, it implicitly includes important design principles, that will be shortly discussed next.\\
First, note that Assumption \ref{Assump:term_constr}.(\ref{Assump:term_constr_1}) can always be verified in the light of the Assumption \ref{eq:Assump_LL}.(\ref{eq:Assump_LL_Schur}) on stability of the transition matrix $A_{\rm\scriptscriptstyle L}$, and establishes a lower bound for the parameter $N_{\rm\scriptscriptstyle L}$. On the other hand, in view of Assumption~\ref{Assump:3}.(\ref{Assump:3_1}), $\beta_{i}$ are full rank and the full rank of matrix $\mathcal{R}_i(N_i)$ is guaranteed by the reachability of the local submodels used at low level control guaranteed by Assumption~\ref{eq:Assump_LL}.(\ref{eq:Assump_LL_Reach}):  therefore Assumption~\ref{Assump:term_constr}.(\ref{Assump:term_constr_rk})
		is fulfilled by taking $N_i$ (and so $N_{\rm\scriptscriptstyle L}$) sufficiently large.\\
Secondly, Assumptions \ref{Assump:term_constr}.(\ref{Assump:term_constr_kappa})-\ref{Assump:term_constr}.(\ref{Assump:term_constr_5}) provide a tradeoff for the selection of parameters $\rho_{\delta\hat{u}_i}$, $i=1,\dots,M$ and $\rho_{\bar{u}}$. In fact\\
- on the one hand, as required by Assumption \ref{Assump:term_constr}.(\ref{Assump:term_constr_5}), and more specifically by equation \eqref{eq:bound u_def}, the input $u(h)$ must be shared between the low-level controllers (with local inputs $\delta {u}_i$, whose maximal amplitude is related to $\rho_{\delta\hat{u}_i}$) and centralized high-level controller (with input $\bar{u}$, whose maximal amplitude is related to $\rho_{\bar{u}}$);\\
- on the other hand, the amplitude of $\delta {u}_i$ must be sufficiently large to compensate for the model inaccuracies, as expressed by Assumptions \ref{Assump:term_constr}.(\ref{Assump:term_constr_kappa}), \ref{Assump:term_constr}.(\ref{Assump:term_constr_4}), and more specifically by equations \eqref{eq:rho_delta_u} and \eqref{Eqn:chidef}.\vspace{1mm}\\
Importantly, note that the constraints \eqref{eq:rho_delta_u} and \eqref{Eqn:chidef} depend upon quantities that are all functions of the number of steps $N_{\rm\scriptscriptstyle L}$, as clarified in the following.\\
- In view of Assumptions~\ref{eq:Assump_LL}.(\ref{eq:Assump_LL_Schur}),~\ref{Assump:3}.(\ref{Assump:3_2}), and~\ref{Assump:3}.(\ref{Assump:3_0}), $\kappa(N_{\rm\scriptscriptstyle L})=\|\sum_{j=0}^{N_{\rm\scriptscriptstyle L}-1}A_{\rm\scriptscriptstyle H}^{j}B_{\rm\scriptscriptstyle H}-G_{\rm\scriptscriptstyle H}(1)-(\beta\sum_{j=0}^{N_{\rm\scriptscriptstyle L}-1}A_{\rm\scriptscriptstyle L}^{j}B_{\rm\scriptscriptstyle L}-\hat{G}_{\rm\scriptscriptstyle L}(1))\|$ and
		$G_{\rm\scriptscriptstyle H}(1)=\sum_{j=0}^{+\infty}A_{\rm\scriptscriptstyle H}^{j}B_{\rm\scriptscriptstyle H}$, $\hat{G}_{\rm\scriptscriptstyle L}(1)=\beta\sum_{j=0}^{+\infty}A_{\rm\scriptscriptstyle L}^{j}B_{\rm\scriptscriptstyle L}$. Therefore
		$\kappa(N_{\rm\scriptscriptstyle L})\leq \|\sum_{j=N_{\rm\scriptscriptstyle L}}^{+\infty}A_{\rm\scriptscriptstyle H}^{j}B_{\rm\scriptscriptstyle H}\|+\|\beta\sum_{j=N_{\rm\scriptscriptstyle L}}^{+\infty}A_{\rm\scriptscriptstyle L}^{j}B_{\rm\scriptscriptstyle L}\|\leq \|A_{\rm\scriptscriptstyle H}^{N_{\rm\scriptscriptstyle L}}\| \|G_{\rm\scriptscriptstyle H}(1)\|+\|A_{\rm\scriptscriptstyle L}^{N_{\rm\scriptscriptstyle L}}\| \|\beta\|\|G_{\rm\scriptscriptstyle L}(1)\|$,
		where ${G}_{{\rm {\rm\scriptscriptstyle L}}}(z)=(zI-A_{{\rm {\rm\scriptscriptstyle L}}})^{-1}B_{{\rm {\rm\scriptscriptstyle L}}}$. Therefore $\kappa(N_{\rm\scriptscriptstyle L})\rightarrow 0$ exponentially as $N_{\rm\scriptscriptstyle L}\rightarrow +\infty$. This shows that also Assumption~\ref{Assump:term_constr}.(\ref{Assump:term_constr_kappa}) can be fulfilled by taking $N_{\rm\scriptscriptstyle L}$ sufficiently large.\\
- Similarly to Proposition 2.3 in \cite{PicassoZhangScat}, for any $i=1,...,M$ it can be proved that
		\begin{equation}
			\lim_{N_{\rm\scriptscriptstyle L}\to+\infty}\|A_{{\rm {\rm\scriptscriptstyle L}}}^{N_{\rm\scriptscriptstyle L}}\|=0,\lim_{N_{\rm\scriptscriptstyle L}\to+\infty}\chi_{i}(N_{\rm\scriptscriptstyle L})=0\label{Eqn:thirdproperty}
		\end{equation}
This proves that also Assumption~\ref{Assump:term_constr}.(\ref{Assump:term_constr_4}) can be fulfilled - i.e., we can increase the maximum amplitude of $\bar{u}$ as much as possible - by taking a sufficiently large value of parameter $N_{\rm\scriptscriptstyle L}$. This, as a byproduct, allows to minimize the conservativity of the overall control scheme, as discussed in the next section.
\subsection{Main results and conservativity of the scheme}
The size of the uncertainty set $\mathcal{W}$ to be considered in the high level design is given by
	\begin{equation}
		\label{eq:W_def}
		\mathcal{W}=\mathcal{B}_{\rho_w}(0)
	\end{equation}
	where $\rho_w=\sum_{j=2}^{N_{\rm\scriptscriptstyle L}}\|\beta F_{\rm\scriptscriptstyle L}^{N_{\rm\scriptscriptstyle L}-j}(A_{\rm\scriptscriptstyle L}-A_{\rm\scriptscriptstyle L}^{\rm\scriptscriptstyle D})\|\rho_{\delta\hat{x}}(j-1)$.\\
The main result can now be stated.
	\begin{theorem}
		\label{theorem:overall_feasibility}
		Under Assumption \ref{Assump:term_constr}, if $x(0)$ is such that the problem \eqref{Eqn:HLoptimiz_1} is feasible at $k=0$ and, for all $i=1,\dots,M$ $$\|x(0)-x_{\rm\scriptscriptstyle S}\|\leq{\frac{(\sqrt{N_i}\underline{\sigma}_{\mathcal{H}_{i}(N_i)}\rho_{\delta \hat{u}_{i}}-\kappa(N_{\rm\scriptscriptstyle L})\rho_{\bar{u}})}{\|{\mathcal{A}}(N_{\rm\scriptscriptstyle L})\|}}:=\lambda_{i}(N_{\rm\scriptscriptstyle L})$$
		then\\
		(i) $\bar{w}(k)\in\mathcal{W}$ and problems \eqref{Eqn:HLoptimiz_1} and \eqref{Eqn:LLoptimiz_1} are feasible for all $k\geq 0$;\\
		(ii) for all $h\geq 0$
		\begin{equation}u(h)\in\bar{u}_{\rm\scriptscriptstyle S}\oplus\mathcal{U}_{\rm\scriptscriptstyle S}\label{eq:bound u}\end{equation}
		(iii) the state of the slow time-scale reduced model $\bar{\Sigma}^{[N_{\rm\scriptscriptstyle L}]}$ enjoys robust convergence properties, i.e.,
		$$\bar{x}^{[N_{\rm\scriptscriptstyle L}]}(k)\rightarrow\bar{x}_{\rm\scriptscriptstyle S}\oplus \mathcal{Z}\text{ as }k\rightarrow+\infty$$
		(iv)  the state of the large scale model $\Sigma$ enjoys robust convergence properties, i.e., for a computable positive constant $\rho_x$
		$$x(k N_{\rm\scriptscriptstyle L})\rightarrow {x}_{\rm\scriptscriptstyle S}\oplus\bigoplus_{h=0}^\infty (F_{\rm\scriptscriptstyle L}^{[N_{\rm\scriptscriptstyle L}]})^h \mathcal{B}_{\rho_x}(0)$$
		(v) we can define (see \eqref{eq:small_gain_final} in Appendix \ref{app:proof}) a function $\sigma(N_{\rm\scriptscriptstyle L})$ of $N_{\rm\scriptscriptstyle L}$ such that, if
		\begin{align}\sigma(N_{\rm\scriptscriptstyle L})<1
			\label{eq:small_gain_final2}\end{align}
		then, as $k\rightarrow \infty$, $x(k N_{\rm\scriptscriptstyle L})\rightarrow {x}_{\rm\scriptscriptstyle S}$.
		\hfill$\square$
	\end{theorem}
	Theorem \ref{theorem:overall_feasibility} establishes three important facts. First, it shows that, if the initial state lies in a suitable set (and if Assumption \ref{Assump:term_constr} holds), the joint feasibility properties of the two control layers can be guaranteed in a recursive fashion. Secondly, it ensures robust convergence of the states of both the reduced-scale and the overall systems to a neighborhood of the corresponding steady-state goals.\\
Finally, if a suitably-defined function $\sigma(N_{\rm\scriptscriptstyle L})$ is smaller than one, then convergence of the state to the goal is ensured. The definition of $\sigma(N_{\rm\scriptscriptstyle L})$ is quite involved and for this reason it is given in Appendix \ref{app:proof}, in particular see equation~\eqref{eq:small_gain_final}. A general remark, however, is due on parameter $\sigma(N_{\rm\scriptscriptstyle L})$: the more the subsystems are interconnected (in a wide sense, regarding both the existence of dependencies between subsystems and their amplitude), the larger $\sigma(N_{\rm\scriptscriptstyle L})$. On the other hand, it is possible to reduce arbitrarily this parameter by increasing the tuning knob $N_{\rm\scriptscriptstyle L}$. This depends of the fact that, as $N_{\rm\scriptscriptstyle L}\rightarrow +\infty$, both $\|{\mathcal{A}}(N_{\rm\scriptscriptstyle L})\|\rightarrow 0$ and $\|{\mathcal{B}}(N_{\rm\scriptscriptstyle L})\|\rightarrow 0$.\\
A further remark is that, similarly to Proposition 2.3 in \cite{PicassoZhangScat}, for any $i=1,...,M$ it can be proved that
$\lim_{N_{\rm\scriptscriptstyle L}\to+\infty}\lambda_{i}(N_{\rm\scriptscriptstyle L})=+\infty$, allowing to increase at will the feasibility region of the low-level problem.\\
From the discussion in Section \ref{subsec:mainass} it has finally become clear that, by tuning the value of the low-level prediction horizon $N_{\rm\scriptscriptstyle L}$, one can reduce at will the values of $\rho_{\delta \hat{u}_i}$, related to the maximum required amplitude of inputs $\delta u_i$. This, from \eqref{Eqn:bound_deltaxhat} and \eqref{eq:W_def}, allows to reduce arbitrarily the high-level disturbance set $\mathcal{W}$. This, in turn, allows to reduce at will the corresponding RPI set $\mathcal{Z}$ and to minimize the conservativity of the present control scheme. We remark that, although a fine tuning of the gain matrices $K_{i}$, $i=1,\dots, M$ and $\bar{K}_{\rm\scriptscriptstyle H}$ can also be beneficial for the reduction of the conservativity of the scheme, the most relevant tuning knob indeed results parameter $N_{\rm \scriptscriptstyle L}$, especially since the dependence upon all other design parameters results rather straightforward and simple.

From the discussion above a further consideration is due. Although the case $N_{\rm \scriptscriptstyle L}\rightarrow+\infty$ allows to verify all the requirements, to obtain the best dynamic performances from the application of our control scheme, it would be beneficial to set $N_{\rm \scriptscriptstyle L}$ to an "average" value, such that the assumptions are verified, but at the same time that guarantees to control the system in a dynamic fashion also at high level. It is nevertheless important to remark that, when $N_{\rm \scriptscriptstyle L}\rightarrow+\infty$, the scheme can be regarded as a two-layer algorithm that at high level consists of a static optimizer based on a simplified system model and, at low level, consists of a dynamic, reactive, decentralized, and multi-rate, optimization-based regulator.
\subsection{Design}
	The implementation of the multilayer algorithm described in the previous section requires a number of off-line computations here listed for the reader's convenience.
	\begin{itemize}
		\item  design of $A_{\rm\scriptscriptstyle H}$, $B_{\rm\scriptscriptstyle H}$, and $\beta_i$, $i=1,...,M$, such that Assumption \ref{Assump:3} is satisfied;
		\item design of $\bar{K}_{\rm\scriptscriptstyle H}$ such that both $F_{\rm\scriptscriptstyle H}=A_{{\rm {\rm\scriptscriptstyle H}}}^{N_{\rm\scriptscriptstyle L}}+B_{{\rm {\rm\scriptscriptstyle H}}}^{{[N_{\rm\scriptscriptstyle L}]}}\bar{K}_{\rm\scriptscriptstyle H}$ and
		$F_{\rm\scriptscriptstyle L}^{[N_{\rm\scriptscriptstyle L}]}=A_{\rm\scriptscriptstyle L}^{N_{\rm\scriptscriptstyle L}}+B_{\rm\scriptscriptstyle L}^{[N_{\rm\scriptscriptstyle L}]}\bar{K}_{\rm\scriptscriptstyle H}\beta$ are Schur stable;
		\item design of $K=\text{diag}(K_1,\dots,K_{\rm\scriptscriptstyle M})$ such that $F_{\rm\scriptscriptstyle L}=A_{\rm\scriptscriptstyle L}+B_{\rm\scriptscriptstyle L}K$ is Schur stable;
		\item computation of $\rho_{\delta \hat{u}_i}$, $\rho_{\bar{u}_i}$ (see the procedure proposed in Appendix \ref{app:rhos}) and of the sets ${\bar{\mathcal{U}}}_{{\rm\scriptscriptstyle S},i}$, $\Delta\hat{\mathcal{U}}_i$;
		\item computation of $\mathcal{W}$  according to \eqref{eq:W_def} and \eqref{Eqn:bound_deltaxhat};
		\item computation of $\bar{\mathcal{X}}_{\rm\scriptscriptstyle F}$, $\mathcal{Z}$, see \cite{RawlingsBook}, and $P_{\rm\scriptscriptstyle H}$ with \eqref{eqn:HL_Lyap}.
	\end{itemize}
	\section{Simulation example and implementation procedures}
	The hierarchical control algorithm described in the previous sections has been used to control the model of the large-scale chemical plant described in~\cite{LML},~\cite{HDMPC}.
\subsection{Description of the plant and linearized model}
\begin{figure}[ht]
		\center
		\includegraphics[width=0.95\columnwidth]{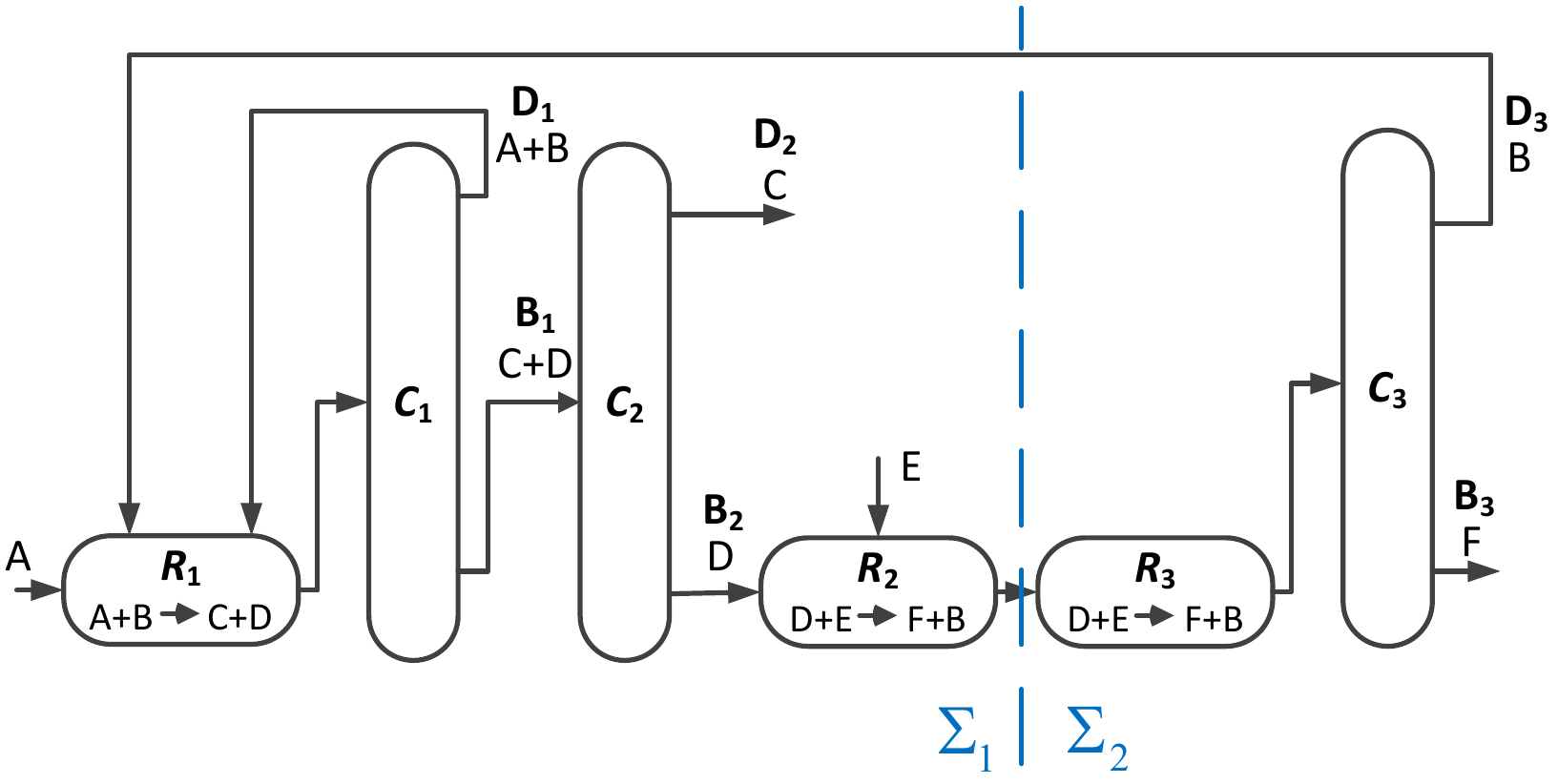}
		\caption{Flow diagram of the chemical plant: for $i=1,2,3$, $R_i$ and $C_i$ are the reactors and distillation columns, while $D_i$ and $B_i$ are the top and bottom products.}
		\label{fig:chemicalplant}
	\end{figure}

The system is composed of three reactors $R_1$, $R_2$, $R_3$, three distillation columns $C_1$, $C_2$, $C_3$, two recycle streams and six chemical components: $A,B,C,D,E,F$. The flow diagram of the system is reported in Figure~\ref{fig:chemicalplant}, where $D_1$, $D_2$, $D_3$ are the top products of the columns, while $B_1$, $B_2$, $B_3$ are their bottom products. The following reactions occur inside the reactors $R_{1}:\,A+B\rightarrow C+D$, $R_{2}:\,D+E\rightarrow F+B$, and $R_{3}:\,D+E\rightarrow F+B$.
	The system has six control variables, namely the refluxes ($v_1,v_3,v_5$) and the vapour ($v_2,v_4,v_6$) flow rates in the columns $C_{1},C_{2},C_{3}$, respectively, and six outputs: the liquid molar fraction ($r_1$) of component $A$ at the top product of $C_{1}$, the liquid molar fraction ($r_2$) of component $D$ at the bottom product of $C_{1}$,  the liquid molar fraction ($r_3$) of component $C$ at the top product of $C_{2}$, the liquid molar fraction ($r_4$) of component $D$ at the bottom product of $C_{2}$, the liquid molar fraction ($r_5$) of component $B$ at the top product of $C_{3}$, and the liquid molar fraction ($r_6$) of component $F$ at the bottom product of $C_{3}$. A detailed description of the model equations and of the model parameters is reported in~\cite{HDMPC}. The considered nominal operating point is characterized by the vector of inputs $v_{nominal}=[330 \quad 410 \quad 283 \quad 385 \quad 141 \quad 282]^{'}$ lb$\,$mol/h, to which it corresponds the vector of outputs $r_{nominal}=[0.942\quad0.552\quad0.827\quad0.941\quad0.705\quad 0.991]^{'}$.  The linearized model at this operating condition is of order $n=192$, and shows strong interactions among the control and controlled variables, see again~\cite{HDMPC}.\\
	In order to apply the hierarchical control structure described in this paper, the system and the corresponding linearized model have been partitioned into two subsystems (i.e. $M=2$). The first one includes reactors $R_{1},R_{2}$ and columns $C_{1},C_{2}$, while the second one is made by $R_{3}$ and $C_{3}$.
    The continuous-time linearized models of the two subsystems have been discretized with
    the algorithm described in~\cite{farina2013block} and basic sampling time $T=0.05$ h and a standard model order reduction procedure has been used to remove the unreachable states of the two linear subsystems. Denoting with $\delta v_i$ and $\delta r_i$ the deviations of the inputs $v_i$ and the outputs $r_i$, respectively, with respect to their nominal values, the first resulting linear reduced order model is of order $n_{1}=25$, with $m_1=4$ inputs, i.e., $u_1=(\delta v_1,\cdots,\delta v_4)$, $p_1=4$ outputs, i.e., $y_1=(\delta r_1,\cdots,\delta r_4)$, and $p_{z1}=3$ coupling outputs, while the second one has $n_2=16$ states, $m_2=2$ inputs, i.e., $u_2=(\delta v_5,\delta v_6)$, $p_2=2$ outputs, i.e., $y_2=(\delta r_5,\delta r_6)$, and $p_{z2}=2$ coupling outputs.
    These two linear subsystems have been used as the linear models described in \eqref{Eqn:LL} for the implementation of the hierarchical control structure. The control variables are limited by
    \begin{equation}\label{Eqn:u_i_exam}
    \|u_{i}-u_{\rm \scriptscriptstyle S,i}\|_{\infty}\leq 100,
    \end{equation}
     for $i=1,2$, where $u_{\rm \scriptscriptstyle S, i}$ is the steady state value corresponding to the output set-point values given by $y_{\rm\scriptscriptstyle S}=10^{-3}\cdot (9.4,\, 5.5, \, 8.3, \, 9.4, \, 7.0, \, 9.9)$.
\subsection{Design and implementation procedures}

    The  procedures used to implement the hierarchical control structure previously described and the computational details    are now listed.\\
\subsubsection{ Tuning the hierarchical control scheme with  $N_{\rm \scriptscriptstyle L}=28$, $N_{\rm \scriptscriptstyle H}=3$}\label{subsec:tuning_1}
\paragraph{\emph{Devising the high-level simplified model and the low-level submodels}}
    \begin{itemize}
    \item The procedure described in Appendix~\ref{app:beta} has been used to compute the matrices
	$\beta_{1}$ and $\beta_{2}$, and the reduced order model \eqref{Eqn:C HL model}
	with order $\bar{n}=6$. The dynamic matrix
	$A_{\rm \scriptscriptstyle H}={\rm diag}(0.972,\,0.984,\,0.969,\,0.969,\,0.874,\,0.869)$ has been computed; its parameters have been selected as the maximum singular values of the reachability matrix of each subsystem previously discretized. The matrix $B_{\rm \scriptscriptstyle H}$ has been computed as described in Appendix \ref{app:beta}. The resulting model has been resampled with $N_{\rm \scriptscriptstyle L}=28$ to obtain the model \eqref{Eqn:C HLS model} to be used at the high level in the slow time scale.
	\item The models \eqref{Eqn:LL dndyn}  have been re-sampled   with $\zeta_1=4$ and $\zeta_2=1$ to obtain the models \eqref{Eqn:LL dndynopt} to be used at the low level in the fast time scale.
\end{itemize}
\paragraph{\emph{Off-line design of the low-level regulators}}
	\begin{itemize}
		\item The low level finite-horizon optimization algorithms described in \eqref{Eqn:LLoptimiz_1} and \eqref{Eqn:LL_cost} have been implemented with
	 state and input penalties $Q_{1}=10^3\cdot\beta_1^{\rm\scriptscriptstyle T}\beta_1$, $Q_{2}=10^4\cdot\beta_2^{\rm\scriptscriptstyle T}\beta_2$,
	$R_{1}=I_{m_{1}}$ and $R_{2}=I_{m_{2}}$.
	
	\item The decentralized feedback gains $K_1$ and $K_2$ guaranteeing that $F_{\rm\scriptscriptstyle L}$ is Schur stable can in principle be computed according to the algorithm described in \cite{GiulioB} and based on the solution of LMI problems. However, in our case, we simply solved two independent LQ problems and we checked that the resulting $F_{\rm\scriptscriptstyle L}$ is Schur stable.
	\item The parameters $\rho_{\delta \hat u_1}$, $\rho_{\delta \hat u_2}$, $\rho_{\bar u_1}$ and $\rho_{\bar u_2}$   have been computed according to \eqref{Eqn:linearoptmiz_u} in Appendix \ref{app:rhos}. Specifically we used the cost function $J_{\rho}=\gamma_1 \mathds{1}_{\rm\scriptscriptstyle 1\times M}\overrightarrow{\rho}_{\delta\hat{u}}-(\overrightarrow{\rho}_{\bar{u}}-\gamma_2\overrightarrow{\rho}_{u})^{'}(\overrightarrow{\rho}_{\bar{u}}-\gamma_2\overrightarrow{\rho}_{u})$, where $\gamma_1$, $\gamma_2$ are  positive constants selected as $\gamma_1=1$ and $\gamma_2=0.3$. Note that the choice of $\gamma_1, \gamma_2$ allows one to modify the feasibility region of the high level optimization problem \eqref{Eqn:HLoptimiz_1} and \eqref{Eqn:HL_cost}; typically setting $\gamma_1=1$, the feasibility properties of \eqref{Eqn:HLoptimiz_1} grow with  $\gamma_2$.
The computed values are  $\rho_{\delta \hat u_1}=145.2$, $\rho_{\delta \hat u_2}=115.7$, $\rho_{\bar u_1}=49.3$, $\rho_{\bar u_2}=25.2$. The input vectors of each subsystem, namely $\bar{u}_i$, $\delta \hat{u}_i$, $i=1,2$, have been limited as follows: $-\rho_{\bar u_i}/\sqrt{m_i}\mathds{1}_{m_i\times 1}\leq\bar u_i-\bar u_{\rm\scriptscriptstyle S, i}\leq\rho_{\bar u_i}/\sqrt{m_i}\mathds{1}_{m_i\times 1}$, $-\rho_{\delta \hat u_i}/\sqrt{m_i}\mathds{1}_{m_i\times 1}\leq\delta \hat u_i\leq\rho_{\delta \hat u_i}/\sqrt{m_i}\mathds{1}_{m_i\times 1}$; and the corresponding sets $\bar{\mathcal{U}}_{\rm\scriptscriptstyle S,i}$, $\Delta\hat{\mathcal{U}}_{\rm\scriptscriptstyle S,i}$, $i=1,2$ have been obtained.

\end{itemize}
  Note that $\rho_{\bar u_1}/\sqrt{m_1}+\rho_{\delta \hat u_1}/\sqrt{m_1}=97.3$ and  $\rho_{\bar u_2}/\sqrt{m_2}+\rho_{\delta \hat u_2}/\sqrt{m_2}=99.6$ and we can write $\bar{\mathcal{U}}_{\rm\scriptscriptstyle S,1}\oplus\Delta\hat{\mathcal{U}}_{\rm\scriptscriptstyle S,1}=\{\bar u_1+\delta \hat u_1|-97.3\cdot\mathds{1}_{m_1\times 1}\leq\bar u_1+\delta \hat u_1-\bar u_{\rm\scriptscriptstyle S, 1}\leq97.3\cdot\mathds{1}_{m_1\times 1}\}$  and $\bar{\mathcal{U}}_{\rm\scriptscriptstyle S,2}\oplus\Delta\hat{\mathcal{U}}_{\rm\scriptscriptstyle S,2}=\{\bar u_2+\delta \hat u_2|-99.6\cdot\mathds{1}_{m_2\times 1}\leq\bar u_1+\delta \hat u_2-\bar u_{\rm\scriptscriptstyle S, 2}\leq99.6\cdot\mathds{1}_{m_2\times 1}\}$. These results show that the size of both the sets $\bar{\mathcal{U}}_{\rm\scriptscriptstyle S,1}\oplus\Delta\hat{\mathcal{U}}_{\rm\scriptscriptstyle S,1}$ and $\bar{\mathcal{U}}_{\rm\scriptscriptstyle S,2}\oplus\Delta\hat{\mathcal{U}}_{\rm\scriptscriptstyle S,2}$ is close to that of the real input constraint in \eqref{Eqn:u_i_exam}.
  In addition, the radius of the sets $\mathcal{B}_{\rho_{\Delta u_i(N_{\rm\scriptscriptstyle L}-1)}}(0)$, $i=1,2$, resulting from the feedback policy in \eqref{Eqn:LL dndyn_sf_deltau} for compensating the couplings terms between subsystems have been computed according to Assumption \ref{Assump:term_constr}.(\ref{Assumption_bound_u_def}) with $\rho_{\Delta u_1(N_{\rm\scriptscriptstyle L}-1)}=4.0$ and $\rho_{\Delta u_2(N_{\rm\scriptscriptstyle L}-1)}=0.37$. Therefore, we can also write
  $\bar{\mathcal{U}}_{\rm\scriptscriptstyle S,1}\oplus\Delta\mathcal{U}_{\rm\scriptscriptstyle S,1}=\{\bar u_1+\delta  u_1|-99.3\cdot\mathds{1}_{m_1\times 1}\leq\bar u_1+\delta u_1-\bar u_{\rm\scriptscriptstyle S, 1}\leq99.3\cdot\mathds{1}_{m_1\times 1}\}$ and $\bar{\mathcal{U}}_{\rm\scriptscriptstyle S,2}\oplus\Delta\mathcal{U}_{\rm\scriptscriptstyle S,2}=\{\bar u_2+\delta u_2|-99.9\cdot\mathds{1}_{m_2\times 1}\leq\bar u_1+\delta u_2-\bar u_{\rm\scriptscriptstyle S, 2}\leq99.9\cdot\mathds{1}_{m_2\times 1}\}$.
   In view of this, the conservativeness of the algorithm related to the computation of the input constraint sets described in Appendix \ref{app:rhos} is small, especially for the second subsystem.
\paragraph{\emph{Off-line design of the high-level regulator}}
	 \begin{itemize}
	 \item  The high-level tube-based robust MPC has been designed  according to the algorithm
	described in \cite{Mayne2005219} with state and input penalties $Q_{\rm \scriptscriptstyle H}=I_{\bar{n}}$ and $R_{\rm \scriptscriptstyle H}=0.1I_{m}$.
	\item The gain matrix $\bar K_{\rm \scriptscriptstyle H}$ guaranteeing that  both $F_{\rm\scriptscriptstyle H}$ and $F_{\rm\scriptscriptstyle L}^{[N_{\rm\scriptscriptstyle L}]}$ are Schur stable can be computed according to the algorithm described in \cite{GiulioB} and based on the solution of LMI problems.  However, in our case, we simply solved  a LQ problem and we checked that the resulting  $F_{\rm\scriptscriptstyle H}$ and $F_{\rm\scriptscriptstyle L}^{[N_{\rm\scriptscriptstyle L}]}$ were Schur stable.
	\item The disturbance set has been obtained according to  \eqref{eq:W_def} and \eqref{Eqn:bound_deltaxhat} with $\rho_w=1.25$, and the RPI set has been computed with the algorithms described at pp. 231-233 of \cite{rawlings2009model}, i.e., $\mathcal{Z}=\{ z|-(2.66,3.61,1.93,2.64,1.30,1.30)\leq z\leq\break(2.66,3.61,1.93,2.64,1.30,1.30)\}$ and $\bar K_{\rm \scriptscriptstyle H}\mathcal Z=\{\bar K_{\rm \scriptscriptstyle H}z|-(1.64,1.63,1.02,1.02,\break0.10,0.10)\leq \bar K_{\rm \scriptscriptstyle H}z\leq(1.64,1.63,1.02,1.02,0.10,0.10)\}$. The terminal penalty $P_{\rm \scriptscriptstyle H}$ is computed with \eqref{eqn:HL_Lyap}, and the terminal set is calculated under nominal input constraints in \eqref{Eqn:HLoptimiz_1},
i.e.,$\bar {\mathcal{X}}_{\rm \scriptscriptstyle F}=\{\bar{x}^{[N_{\rm \scriptscriptstyle L}],o}|(\bar{x}^{[N_{\rm \scriptscriptstyle L}],o}-\bar{x}_{\rm\scriptscriptstyle S})^{T}P_{\rm \scriptscriptstyle H}(\bar{x}^{[N_{\rm \scriptscriptstyle L}],o}-\bar{x}_{\rm\scriptscriptstyle S})\leq0.997\}$, see \cite{hu2008model}.\\
\end{itemize}
The values taken by $\rho_w$, $\mathcal{Z}$ and $\bar K_{\rm \scriptscriptstyle H}\mathcal Z$ reveal that the feasibility region of high-level regulator might be slightly reduced compared to the one of  stabilizing MPC due to the use of tightened input constraint set, i.e., ${\bar{u}_{\rm\scriptscriptstyle S}\oplus\bar{\mathcal{U}_{\rm\scriptscriptstyle S}}\ominus \bar{K}_{\rm\scriptscriptstyle H}\mathcal{Z}}$ in optimization problem \eqref{Eqn:HLoptimiz_1} and the computation formulas of the disturbance set in \eqref{eq:W_def} and \eqref{Eqn:bound_deltaxhat}, but can be enlarged by increasing the tuning knobs $\rho_{\bar u_1}$ and $\rho_{\bar u_2}$.
\subsubsection{Tuning the hierarchical control scheme with $N_{\rm \scriptscriptstyle L}=84$, $N_{\rm \scriptscriptstyle H}=1$}
Along the same line with the previous Section \ref{subsec:tuning_1}, the computational results corresponding to tuning knob $N_{\rm \scriptscriptstyle L}=84$ and $N_{\rm \scriptscriptstyle H}=1$ are also listed here.
\begin{itemize}
	\item
The following parameters $\rho_{\delta \hat u_1}$, $\rho_{\delta \hat u_2}$, $\rho_{\bar u_1}$ and $\rho_{\bar u_2}$ have been computed: $\rho_{\delta \hat u_1}=125.6$, $\rho_{\delta \hat u_2}=98.3$, $\rho_{\bar u_1}=59.5$, $\rho_{\bar u_2}=41.9$. The amplitude of both  $\rho_{\delta \hat u_1}$ and $\rho_{\delta \hat u_2}$ ($\rho_{\bar u_1}$ and $\rho_{\bar u_2}$)  is smaller (larger) than that with tuning knob $N_{\rm \scriptscriptstyle L}=28$, $N_{\rm \scriptscriptstyle H}=3$.
\item The sets $\bar{\mathcal{U}}_{\rm\scriptscriptstyle S,1}\oplus\Delta\hat{\mathcal{U}}_{\rm\scriptscriptstyle S,1}=\{\bar u_1+\delta \hat u_1|-92.6\cdot\mathds{1}_{m_1\times 1}\leq\bar u_1+\delta \hat u_1-\bar u_{\rm\scriptscriptstyle S, 1}\leq92.6\cdot\mathds{1}_{m_1\times 1}\}$  and $\bar{\mathcal{U}}_{\rm\scriptscriptstyle S,2}\oplus\Delta\hat{\mathcal{U}}_{\rm\scriptscriptstyle S,2}=\{\bar u_2+\delta \hat u_2|-99.1\cdot\mathds{1}_{m_2\times 1}\leq\bar u_1+\delta \hat u_2-\bar u_{\rm\scriptscriptstyle S, 2}\leq99.1\cdot\mathds{1}_{m_2\times 1}\}$. These results show that the size of both the sets $\bar{\mathcal{U}}_{\rm\scriptscriptstyle S,1}\oplus\Delta\hat{\mathcal{U}}_{\rm\scriptscriptstyle S,1}$ and $\bar{\mathcal{U}}_{\rm\scriptscriptstyle S,2}\oplus\Delta\hat{\mathcal{U}}_{\rm\scriptscriptstyle S,2}$ has been reduced slightly compared to that with tuning knob $N_{\rm \scriptscriptstyle L}=28$, $N_{\rm \scriptscriptstyle H}=3$, however, is still close to that of the real input constraint in \eqref{Eqn:u_i_exam}.
In addition, the radius of the sets $\mathcal{B}_{\rho_{\Delta u_i(N_{\rm\scriptscriptstyle L}-1)}}(0)$, $i=1,2$, have been computed  with $\rho_{\Delta u_1(N_{\rm\scriptscriptstyle L}-1)}=11.3$ and $\rho_{\Delta u_2(N_{\rm\scriptscriptstyle L}-1)}=0.91$. Therefore, we can also write
	$\bar{\mathcal{U}}_{\rm\scriptscriptstyle S,1}\oplus\Delta\mathcal{U}_{\rm\scriptscriptstyle S,1}=\{\bar u_1+\delta  u_1|-98.2\cdot\mathds{1}_{m_1\times 1}\leq\bar u_1+\delta u_1-\bar u_{\rm\scriptscriptstyle S, 1}\leq98.2\cdot\mathds{1}_{m_1\times 1}\}$ and $\bar{\mathcal{U}}_{\rm\scriptscriptstyle S,2}\oplus\Delta\mathcal{U}_{\rm\scriptscriptstyle S,2}=\{\bar u_2+\delta u_2|-99.8\cdot\mathds{1}_{m_2\times 1}\leq\bar u_1+\delta u_2-\bar u_{\rm\scriptscriptstyle S, 2}\leq99.8\cdot\mathds{1}_{m_2\times 1}\}$.
	In view of this, the conservativity of the algorithm related to the computation of the input constraint sets described in Appendix \ref{app:rhos} is still small, especially for the second subsystem.
	\item The disturbance set has been obtained  with $\rho_w=1.56$, and the RPI set has been computed i.e., $\mathcal{Z}=\{ z|-(1.83,2.11,1.69,1.82,1.58,1.58)\leq z\leq\break(1.83,2.11,1.69,1.82,1.58,1.58)\}$  and $\bar K_{\rm \scriptscriptstyle H}\mathcal Z=\{\bar K_{\rm \scriptscriptstyle H}z|-(0.37,0.29,0.17,0.18,\break0.01,0.01)\leq \bar K_{\rm \scriptscriptstyle H}z\leq(0.37,0.29,0.17,0.18,0.01,0.01)\}$. The terminal penalty $P_{\rm \scriptscriptstyle H}$ is computed with \eqref{eqn:HL_Lyap},  and the terminal set is calculated under nominal  input constraints in \eqref{Eqn:HLoptimiz_1},
	i.e.,$\bar {\mathcal{X}}_{\rm \scriptscriptstyle F}=\{\bar{x}^{[N_{\rm \scriptscriptstyle L}],o}|(\bar{x}^{[N_{\rm \scriptscriptstyle L}],o}-\bar{x}_{\rm\scriptscriptstyle S})^{T}P_{\rm \scriptscriptstyle H}(\bar{x}^{[N_{\rm \scriptscriptstyle L}],o}-\bar{x}_{\rm\scriptscriptstyle S})\leq0.98\}$.
	\end{itemize}
	
Note that, the size of both the sets $\mathcal{Z}$ and $\bar K_{\rm \scriptscriptstyle H}\mathcal Z$ is smaller than that with the tuning $N_{\rm \scriptscriptstyle L}=28$, $N_{\rm \scriptscriptstyle H}=3$. The amplitude of $\rho_{\bar u_1}$ and $\rho_{\bar u_2}$ has been enlarged and the feasibility region of the high-level regulator can be increased significantly.
\subsubsection{On-line implementation}
The on-line implementation proceduce is depicted in Algorithm \ref{Atm:1}.
\begin{algorithm}
	initialization\;\\
	\While{ for any integer $k\geq0$}{
		(1) set $\bar{x}^{[N_{\rm\scriptscriptstyle L}]}(k)=\beta x(kN_{\rm\scriptscriptstyle L})$, $\hat{x}(kN_{\rm\scriptscriptstyle L})=x(kN_{\rm\scriptscriptstyle L})$,  $\delta\hat{x}(kN_{\rm\scriptscriptstyle L})=0$ and $\delta x(kN_{\rm\scriptscriptstyle L})=0$\;\\
		(2) compute the high-level input $\bar{u}^{{[N_{\rm\scriptscriptstyle L}]},o}(k|k)$ by solving the
		high-level optimization problem \eqref{Eqn:HLoptimiz_1} and \eqref{Eqn:HL_cost}, and compute
		$\bar{u}^{[N_{\rm\scriptscriptstyle L}]}(k)$ by equation \eqref{eqn:HL_control_input}\;
		
		(3) update $\bar{x}^{[N_{\rm\scriptscriptstyle L}]}(k+1|k)$ and $\bar{x}^{{[N_{\rm\scriptscriptstyle L}]},o}(k+1|k)$
		
		(4) generate $\hat{x}_{i}(h)$ of system \eqref{Eqn:LL sdyn} in an open-loop
		fashion for $N_{\rm\scriptscriptstyle L}$ fast time steps with high-level control action $\bar{u}^{[N_{\rm\scriptscriptstyle L}]}(\lfloor h/N_{\rm\scriptscriptstyle L}\rfloor)$, for all $h\in[kN_{\rm\scriptscriptstyle L},kN_{\rm\scriptscriptstyle L}+N_{\rm\scriptscriptstyle L})$
		
		(5) compute terminal constraints \eqref{eq:terminaln} for low-level regulators
		
		(6) compute $\overrightarrow{\delta \hat{u}^{[\zeta_i]}_i} (kN_i:(k+1)N_i-1|kN_{\rm\scriptscriptstyle L})$ by
		solving the finite-horizon optimization problem \eqref{Eqn:LLoptimiz_1} and \eqref{Eqn:LL_cost} for $i=1,2$
		
		(7) generate $\delta\hat{u}_{i}(h)$ in \eqref{llcontr_2} with $\delta\hat{u}_{i}^{[\zeta_i]}(\lfloor h/\zeta_i\rfloor|kN_{\rm\scriptscriptstyle L})$ and compute  $\delta\hat{x}_{i}(h)$ in \eqref{Eqn:LL dndyn} with  $\delta\hat{u}_{i}(h)$ for $i=1,2$ and for all $h\in[kN_{\rm\scriptscriptstyle L},kN_{\rm\scriptscriptstyle L}+N_{\rm\scriptscriptstyle L})$		\\
		 \For{$h\leftarrow kN_{\rm\scriptscriptstyle L}$ \KwTo $kN_{\rm\scriptscriptstyle L}+N_{\rm\scriptscriptstyle L}-1$}{

		(f1) compute $\delta u_{i}(h)$ with \eqref{llcontr_1}
		and compute  $u_{i}(h)=\bar{u}_i^{[N_{\rm\scriptscriptstyle L}]}(\lfloor h/N_{\rm\scriptscriptstyle L}\rfloor)+\delta u_{i}(h)$ for $i=1,2$

		(f2) update $\delta x(h+1)$ and $x(h+1)$
	}
	(8) k$\leftarrow$k+1
	}
	\caption{On-line implementation}\label{Atm:1}
\end{algorithm}
	
	\begin{table}[h!tb]
		\centering \caption{On-line computation time comparison}
		\label{tab:Tab_com}
		\vskip 0.2cm
		\scalebox{0.6}{
			\begin{tabular}{|c|c|c|c|c|}
				\hline
				\multicolumn{3}{|c|}{Approach}  & Optimization activated at &  Computation time (s)\\
				\hline
				{\multirow{3}{*}{Proposed one with $N_{\rm \scriptscriptstyle L}=28$} }&
				 \multicolumn{2}{c|}{HL regulator}&  $h=kN_{\rm\scriptscriptstyle L}$ & 1.1 \\\cline{2-5}
				&\multirow{2}{*}{LL regulator} &$1^{st}$ one&$h=kN_{\rm\scriptscriptstyle L}$ & 0.94 \\ \cline{3-5}
				&&$2^{nd}$ one& $h=kN_{\rm\scriptscriptstyle L}$& 0.45 \\ \hline
				{\multirow{3}{*}{Proposed one with $N_{\rm \scriptscriptstyle L}=84$} }&
				 \multicolumn{2}{c|}{HL regulator}&  $h=kN_{\rm\scriptscriptstyle L}$ & 1.07 \\\cline{2-5}
				&\multirow{2}{*}{LL regulator} &$1^{st}$ one& $h=kN_{\rm\scriptscriptstyle L}$ & 2.16 \\ \cline{3-5}
				&&$2^{nd}$ one& $h=kN_{\rm\scriptscriptstyle L}$& 3.64 \\ \hline
				 \multicolumn{3}{|c|}{Centralized stabilizing MPC} & each fast time instant $h$ &21.9 \\
				\hline
			\end{tabular}
		}
	\end{table}
	
%

\subsection{Simulation results: application to the linearized model}
\begin{figure}[ht]
	\center
	\includegraphics[width=0.95\columnwidth]{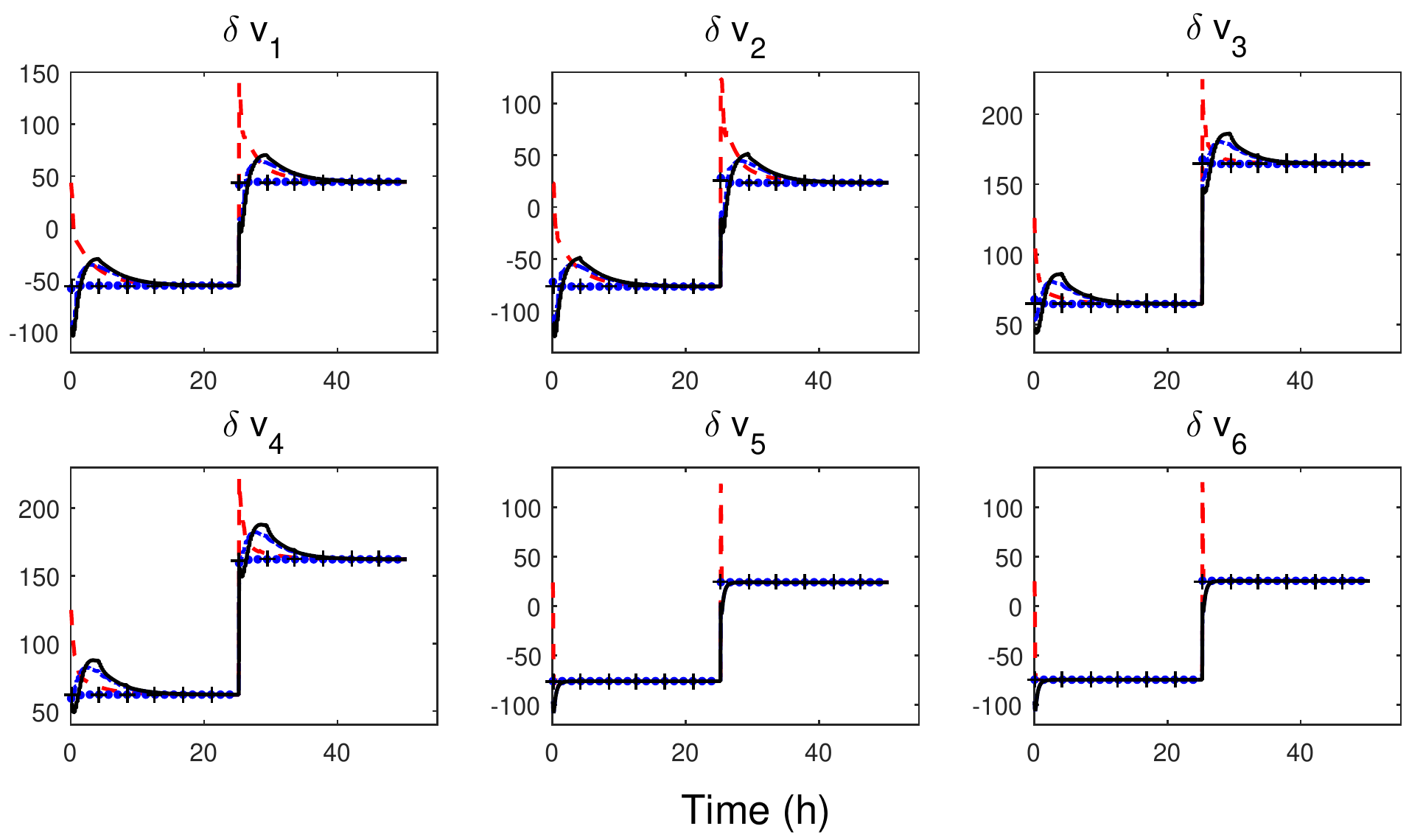}
	\caption{Control variables of the controlled linear system: black $+$ markers (blue $\cdot$ markers) are the values of the inputs computed at the high level by the two-layer scheme with $N_{\rm \scriptscriptstyle L}=84$ and $N_{\rm \scriptscriptstyle H}=1$ (with $N_{\rm \scriptscriptstyle L}=28$ and $N_{\rm \scriptscriptstyle H}=3$), black continuous lines (blue dot-dashed lines) are the values of the overall control actions computed by the two-layer scheme with $N_{\rm \scriptscriptstyle L}=84$ and $N_{\rm \scriptscriptstyle H}=1$ (with $N_{\rm \scriptscriptstyle L}=28$ and $N_{\rm \scriptscriptstyle H}=3$), while red dashed lines are the values of the control variables computed by the centralized scheme.}
	\label{fig:INP}
\end{figure}

\begin{figure}[ht]
		\center
		\includegraphics[width=0.95\columnwidth]{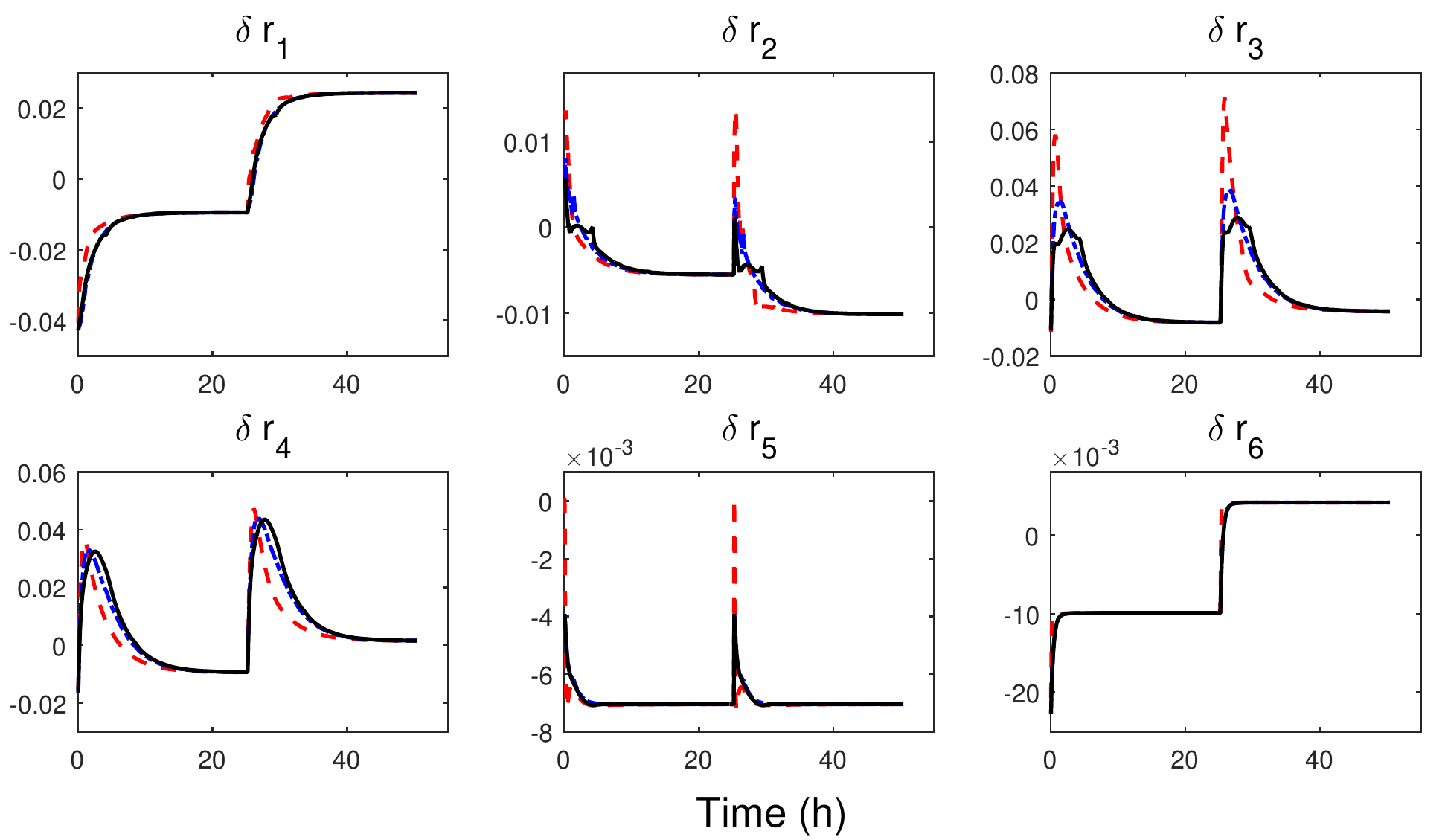}
		\caption{Outputs of the controlled linear system:  black continuous lines (blue dot-dashed lines) are the values of the outputs computed by the two-layer scheme with $N_{\rm \scriptscriptstyle L}=84$ and $N_{\rm \scriptscriptstyle H}=1$ (with $N_{\rm \scriptscriptstyle L}=28$ and $N_{\rm \scriptscriptstyle H}=3$), while red dashed lines are the values of the outputs computed by the centralized scheme.}
		\label{fig:ST1}
	\end{figure}

	The overall control actions computed by the high and low level controllers
	have been applied to the linear system  at each fast sampling time. The output reference values for the linear  system have been initially maintained at the nominal setpoints $y_{\rm\scriptscriptstyle S}=10^{-3}\cdot (9.4,\, 5.5, \, 8.3, \, 9.4, \, 7.0, \, 9.9)$; then, at time $t=25.2$ h, they have been set equal to
	$10^{-2}\cdot(2.43,\,-1.01,\,-0.43,\,0.15,\,-0.70,\,0.41)$. For comparison, a centralized state-feedback stabilizing MPC has been designed with an auxiliary state-feedback control law computed with LQ control, state  penalty matrix $Q=\text{diag}(Q_1,\,Q_2)$, input penalty $R=I_m$ and prediction horizon $N=N_{\rm \scriptscriptstyle H}\cdot N_{\rm \scriptscriptstyle L}=84$. The terminal set has been chosen as $\mathcal{X}_{\rm \scriptscriptstyle F}=\{x|(x-x_{\rm\scriptscriptstyle S})^{T}P(x-x_{\rm\scriptscriptstyle S})\leq1\}$ where the terminal penalty $P$ has been taken as the steady state solution of the Riccati equation according to the infinite horizon control problem with $Q$, $R$. All the simulation tests have been implemented
	within MATLAB Yalmip and MPT toolbox, see \cite{yalmip} and \cite{MPT3}, in a PC with Intel Core i5-4200U
	2.30 GHz and with Windows 10 operating system. The SDPT3 solver has been used for the implementation of the centralized MPC and of the high-level regulator of the proposed approach, while the Matlab QUADPROG solver has been used for the low-level optimization problems. The detailed on-line computational time required for each controller is reported in Table \ref{tab:Tab_com}. {\color{black}This table shows that the on-line computational time of the proposed hierarchical approach for each interval $[kN_{\rm \scriptscriptstyle L},\,kN_{\rm \scriptscriptstyle L}+N_{\rm \scriptscriptstyle L})$ with $N_{\rm \scriptscriptstyle L}=28$ and $N_{\rm \scriptscriptstyle H}=3$ ($N_{\rm \scriptscriptstyle L}=84$ and $N_{\rm \scriptscriptstyle H}=1$), i.e., $2.49$ s ($6.87$ s), is reduced dramatically compared to that of the centralized MPC, i.e., $21.9\cdot N_{\rm \scriptscriptstyle L} $ s}.\\
The evolution of the output and control variables of the controlled linear system is reported
	in Figures \ref{fig:INP}-\ref{fig:ST1} which show that, after an initial transient, inputs and outputs return to their nominal values until the change of the reference occurs, when both the centralized and the two-layer control systems properly react to bring the controlled variables to their reference values, {\color{black}and the performance of the proposed two-layer approach with both tuning cases $N_{\rm \scriptscriptstyle L}=28$, $N_{\rm \scriptscriptstyle H}=3$ and  $N_{\rm \scriptscriptstyle L}=84$, $N_{\rm \scriptscriptstyle H}=1$ is close to that of centralized stabilizing MPC}.

	\begin{figure}[ht]
		\center
		\includegraphics[width=0.95\columnwidth]{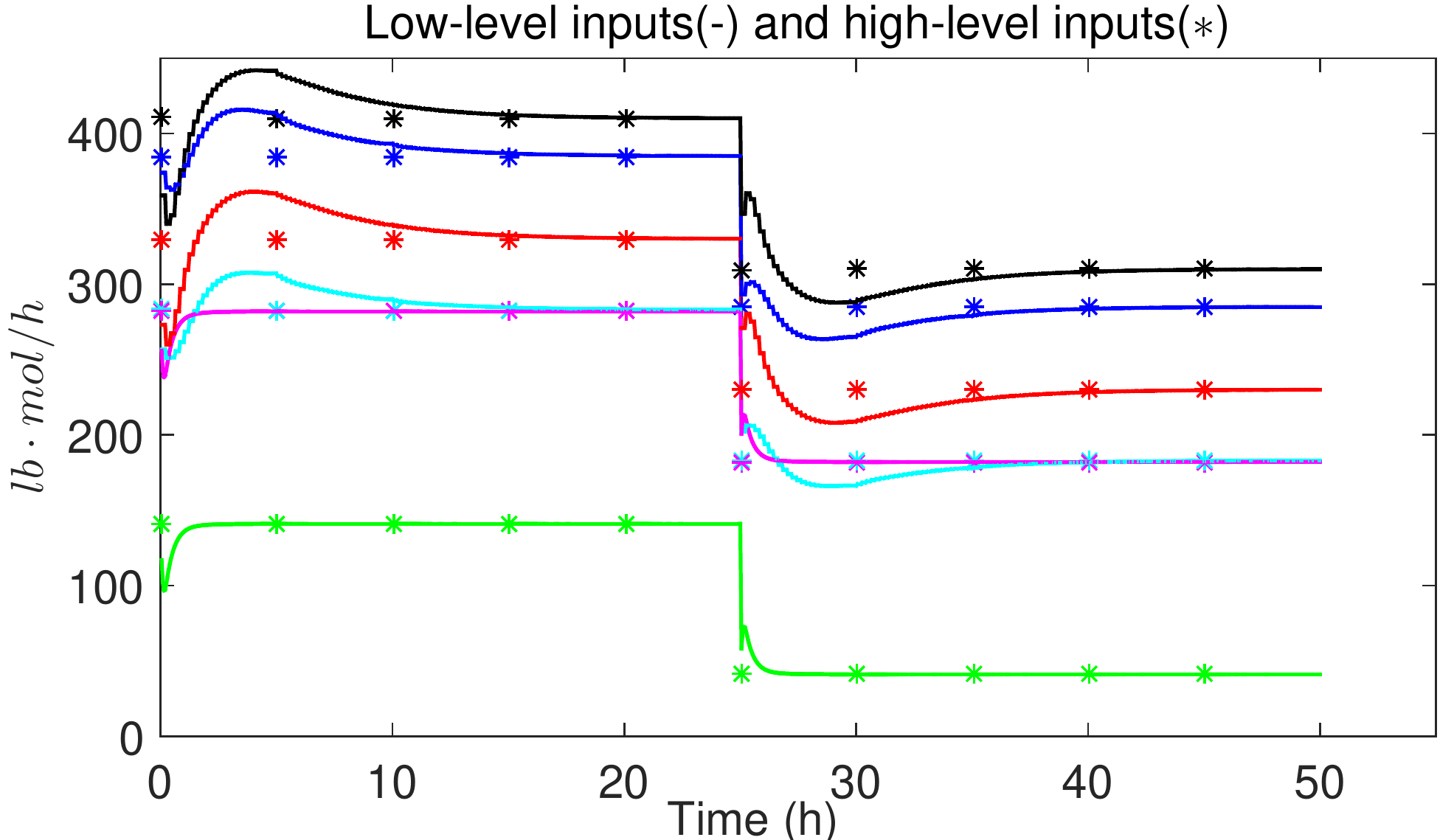}
		\caption{Control variables of the controlled nonlinear system: $\ast$ markers are the values of the control variables computed at the high level in the long sampling time, i.e., $(\bar u_1,\bar u_2)$, (red, black, cyan, blue, green and magenta), while continuous lines are the values of the overall control actions in the multi rate, i.e., $( u_1, u_2)$, (red, black, cyan, blue, green and magenta).}
		\label{fig:INP_n}
	\end{figure}
	\begin{figure}[ht]
		\center
		\includegraphics[width=0.95\columnwidth]{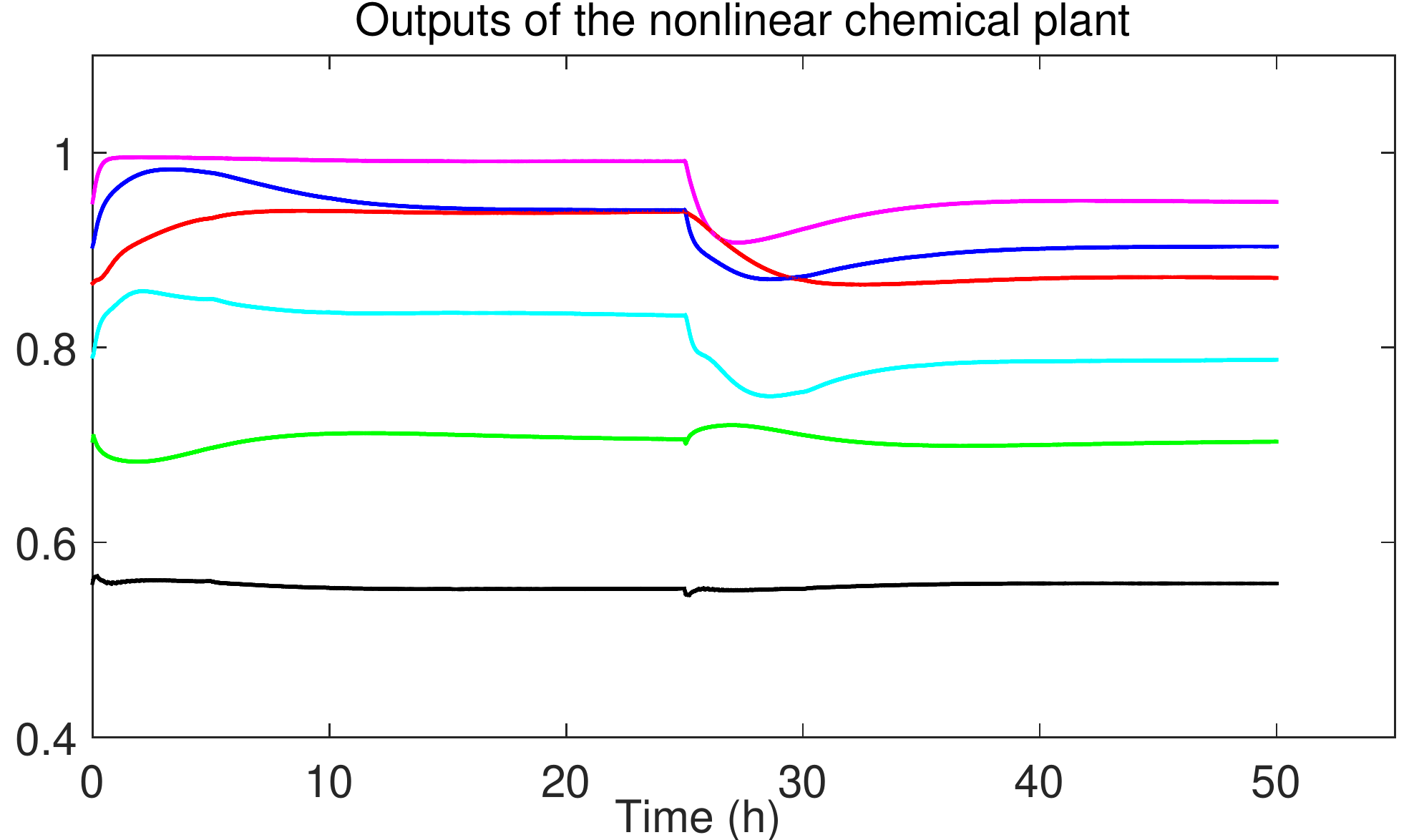}
		\caption{Outputs of the controlled nonlinear system, i.e., $(r_1,\cdots, r_6)$, (red, black, cyan, blue, green and magenta).}
		\label{fig:ST1_n}
	\end{figure}
\subsection{Simulation results: application to the nonlinear system}
The two-layer control structure  has also been applied to the nonlinear chemical plant with $N_{\rm \scriptscriptstyle L}=84$ and $N_{\rm \scriptscriptstyle H}=1$. Since in a realistic scenario the state is unmeasurable, the distributed Kalman filter described in~\cite{farina2015} has been used. The covariances of the noises acting on the states have been set equal to $\hat{Q}_{1}=0.1I_{n_{1}}$, $\hat{Q}_{2}=0.1I_{n_{2}}$,  while the covariances of the output noises have been chosen as $\hat{R}_{1}=0.01I_{p_{1}}$, $\hat{R}_{2}=0.01I_{p_{2}}$. Finally, the covariances of the initial state estimates have been selected as $P_1(0)=0.01I_{n_{1}}$, $P_2(0)=0.01I_{n_{2}}$.\\
Starting from the nominal operating conditions, the overall control actions computed by high and low level controllers
have been applied to the original nonlinear system at each fast time instant. The output reference values for the nonlinear system have been initially maintained at the nominal point $r_{nominal}$; then, at time $t=25.2$ h, they have been set equal to
$[0.866,\,0.558,\,0.791,\,0.903,\,0.704,\,0.948]$. The evolution
of the output and control variables of the controlled nonlinear system are reported
in Figure \ref{fig:INP_n}-\ref{fig:ST1_n}. These figures show that, after an initial transient due to the state filter, inputs and outputs return to their nominal values until the change of the reference occurs, when the two-layer control system properly reacts to bring the controlled variables to their reference values.
	\section{Extensions and conclusions}
	In this paper a two-layer control scheme for systems made by interconnected subsystems has been presented. The algorithm is based on the solution, at the two layers, of MPC problems of reduced size and allows for a multirate implementation, suitable to deal with systems characterized by significantly different dynamics. Its main properties of recursive feasibility and convergence have been established and its performances have been tested in a nontrivial simulation example.\\
The main rationale of the proposed control architecture is grounded on the use, at the high level, of a reduced and slow-timescale model for centralized control. At the low-level, each subsystem is endowed of a local controller and is in charge of both compensating for the model inaccuracies introduced at the high level and dealing with the distributed nature of the system.\\
At high level, in this paper we use a tracking control algorithm based on robust tube-based MPC. This choice, however, is somehow arbitrary, since many alternative (robust) control solutions can be used instead, e.g., the offset-free robust MPC scheme proposed in \cite{Betti13}.\\
The reference signals are here assumed to be previously computed, for instance by an additional Real Time Optimization layer (RTO), e.g., based on the current and predicted external conditions, such as prices of energy or costs of row materials, see \cite{Cutler1983,Forbes1996,Adetola,Kadam,Kadam1,Engell2007,Darby2011,Wurth2011}. It is important to remark that RTO-based structures must be properly designed to guarantee the compatibility of the models used at the two layers, see \cite{Darby2011,Wurth2011,Engell2007}. Also, dynamic RTO structures may be prone to stability issues, as noted in \cite{Ellis2014}.\\
To overcome problems related to the use of RTO to generate reference signals, a robust economic MPC approach can be used, see \cite{bayer2014tube}, for the design of a high level regulator which, at the same time, computes the optimal reference values for the controlled outputs.
This, in our opinion, would not entail significant differences in the algorithm implementation and in the theoretical results. Future work will be devoted to this extension.
%
%
	%
	%
	
	\section{Appendix}
	%
	%
	\subsection{Construction of $\beta_{i}$  and of the reduced order model}
	\label{app:beta}
	A constructive procedure for the computation of the matrices $\beta_{i}$, $i=1,\dots,\,M$, and the reduced order model satisfying Assumption~\ref{Assump:3} is listed here following the same line as in \cite{PicassoZhangScat}. Note however that in \cite{PicassoZhangScat} the case of dynamically decoupled subsystems was considered, and full system reduction actually could be carried out subsystem-by-subsystem. On the contrary, in this case, system reduction must be performed at a full system level and structural additional constraints must be satisfied, i.e. the block-diagonality of matrix $\beta$. In view of the presence of these constraints, the sufficient and necessary condition given in Proposition 1 in \cite{PicassoZhangScat} for guaranteeing the fulfillment of Assumption~\ref{Assump:3} (i.e., that $\bar{n}\geq p+$dim$($Im$G_{\rm\scriptscriptstyle L}^x\cap$Ker$C_{\rm\scriptscriptstyle L})$) is, in our case, only necessary.\\
	Here we now describe a (possibly conservative) procedure:
	\begin{description}
		\item [{a}] find a subspace ${\rm Ker}\,\beta_{i}$ of dimension $n_{i}-\bar{n}_{i}$
		so that ${\rm Ker}\,\beta_{i}\subseteq {\rm Ker}\,C_{{\rm\scriptscriptstyle L}}^{ii}$
		and ${\mathcal{Z}_{ii}}\cap {\rm Ker}\,\beta_{i}=\{0\}$, where $\mathcal{Z}_{ii}={\rm Im}\, \tilde{G}_{\rm\scriptscriptstyle L}^{ii}(1) \cap{\rm Ker}\,C_{\rm\scriptscriptstyle L}^{ii}$, and $ \tilde{G}_{\rm\scriptscriptstyle L}^{ii}(z)=(zI-A_{\rm\scriptscriptstyle L}^{ii})^{-1}\begin{bmatrix}B_{\rm\scriptscriptstyle L}^{ii}&E_i\end{bmatrix}$;
		\item [{b}] let $\{\kappa_{1},\dots,\kappa_{n_{i}-\bar{n}_{i}}\}$ be a
		set of independent vectors in $\mathcal{K}_{\beta,i}=\rm Ker\,\beta_{i}$ and complete this
		set to a basis ${\mathcal{B}}_i=\{v_{1},\dots,v_{\bar{n}_{i}},\kappa_{1},\dots,\kappa_{n_{i}-\bar{n}_{i}}\}$
		of the whole space ${\mathbb{R}}^{n_{i}}$;
		\item [{c}]  let $\{r_{1},\dots,r_{\bar{n}_{i}}\}$ be a basis of ${\mathbb{R}}^{\bar{n}_{i}}$,
		define
		\[
		\hat{\beta_{i}}=\big[\,r_{1}\,|\,\cdots\,|\,r_{\bar{n}_{i}}\,|\,0\,|\,\cdots\,|\,0\,\big]
		\]
		and $T_{{\rm {\scriptscriptstyle L}}}$ be the matrix whose columns
		are the vectors in ${\mathcal{B}}_i$, then $\beta_{i}=\hat{\beta}_{i}T_{{\rm {L}}}^{-1}$;
		\item[{d}] define collective matrix $\beta=\rm{diag}(\beta_1,\dots,\,\beta_{\rm\scriptscriptstyle M})$;
		\item[{e}] choose matrix $A_{\rm\scriptscriptstyle H}$ being Schur stable, and let
		\begin{equation*}
			B_{\rm\scriptscriptstyle H}=(I-A_{\rm\scriptscriptstyle H})\hat{G}_{\rm\scriptscriptstyle L}(1)
		\end{equation*}
		where the suitable choices for $A_{\rm\scriptscriptstyle H}$ are those modeling the dominant dynamics of the low-level collective model.
		
	\end{description}
	Steps \textbf{a}-\textbf{c} imply that Assumption 2.(\ref{Assump:3_1}) is fulfilled, while step \textbf{e} guarantees that Assumption 2.(\ref{Assump:3_0}) and 2.\ref{Assump:3_2} are satisfied.\\
	Note that a less conservative choice can be taken, i.e., defining, in step \textbf{a}, $\tilde{G}_{\rm\scriptscriptstyle L}^{ii}(z)=(zI-A_{\rm\scriptscriptstyle L}^{ii})^{-1}B_{\rm\scriptscriptstyle L}^{ii}$. However this choice does not guarantee a-priori that the property $\mathcal{Z}\cap\mathcal{K}_\beta=\{0\}$, where $\mathcal{Z}=$Im$G_{\rm\scriptscriptstyle L}^x \cap $Ker$C_{\rm\scriptscriptstyle L}$, $G_{\rm\scriptscriptstyle L}^x=(I-A_{\rm\scriptscriptstyle L})^{-1}B_{\rm\scriptscriptstyle L}$, and $\mathcal{K}_\beta=\prod_{i=1}^M\mathcal{K}_\beta^{i}$. This must be verified after the reduction phase has been carried out.
	\subsection{Computation of the input constraint sets}
	\label{app:rhos}
	In the scheme proposed in this paper, the dimensions of the input constraint sets $\bar{\mathcal{U}}_{{\rm\scriptscriptstyle S},i}$ and $\Delta\hat{\mathcal{U}}_i$ are key tuning knobs, which must be selected in order to satisfy, at the same time, the inequalities~\eqref{eq:rho_delta_u}, for all $i=1,\dots,M$, and~\eqref{eq:bound     u_def}. To address the design issue, in this appendix we propose a simple and lightweight algorithm based on a linear program. As a simplifying assumption, we set $\Delta\hat{\mathcal{U}}_i=\mathcal{B}_{\rho_{\delta \hat{u}_i}}(0)$ and $\bar{\mathcal{U}}_{{\rm\scriptscriptstyle S},i}=\mathcal{B}_{\rho_{\bar{u}_i}}(0)$. Under this assumption, the tuning knobs are the vectors $\overrightarrow{\rho}_{\delta\hat{u}}=(\rho_{\delta\hat{u}_1},\dots,\rho_{\delta\hat{u}_{\rm\scriptscriptstyle M}})$ and $\overrightarrow{\rho}_{\bar{u}}=(\rho_{\bar{u}_1},\dots,\rho_{\bar{u}_{\rm\scriptscriptstyle M}})$. Note that, in case of need, such assumption can be relaxed, at the price of a slightly different definition of the inequalities below.
	
	First consider inequality~\eqref{eq:rho_delta_u}, to be verified for all $i=1,\dots,M$. Here the constant $\rho_{\bar{u}}$ appears, defined in such a way that $\bar{\mathcal{U}}_{{\rm\scriptscriptstyle S}}=\prod_{i=1}^M\mathcal{B}_{\rho_{\bar{u}_i}}(0)\subseteq \mathcal{B}_{\rho_{\bar{u}}}(0)$. We can define, for example, $\rho_{\bar{u}}=\sqrt{\sum_{i=1}^M \rho_{\bar{u}_i}^2}\leq \sum_{i=1}^M \rho_{\bar{u}_i}$. Therefore, to fulfill~\eqref{eq:rho_delta_u} it is sufficient to verify the following matrix inequality
	\begin{equation}\label{eq:rho_delta_u_2}
		\overrightarrow{\rho}_{\delta\hat{u}}>\kappa(N_{\rm\scriptscriptstyle L})
		{\rm{diag}}(\frac{1}{\sqrt{N_1}\underline{\sigma}_{\mathcal{H}_{1}(N_1)}},\dots,\frac{1}{\sqrt{N_{\rm\scriptscriptstyle M}}\underline{\sigma}_{\mathcal{H}_{ \rm\scriptscriptstyle M}(N_{\rm\scriptscriptstyle M})}})\mathds{1}_{\rm\scriptscriptstyle M\times M}\overrightarrow{\rho}_{\bar{u}}
	\end{equation}
	where $\mathds{1}_{\rm\scriptscriptstyle M\times M}$ is the $M\times M$ matrix whose entries are all equal to~$1$.
	The second main inclusion to be fulfilled is~\eqref{eq:bound     u_def}, which is verified if, for all $i=1,\dots,M$,
	\begin{equation}\label{eq:bound     u_i}
		\Delta \bar{\mathcal{U}}_i\oplus\bar{\mathcal{U}}_{{\rm\scriptscriptstyle S},i} \subseteq \mathcal{U}_{{\rm\scriptscriptstyle S},i}
	\end{equation}
	By definition, $\Delta \bar{\mathcal{U}}_i=\Delta\hat{\mathcal{U}}_i\oplus\mathcal{B}_{\rho_{\Delta u_i(N_{\rm\scriptscriptstyle L}-1)}}(0)$, where
	\begin{equation}\label{eq:lambda_ij}
		\begin{array}{cll}
			& \rho_{\Delta{u}_i}(N_{\rm\scriptscriptstyle L}-1)=\\
			&=\sum_{r=2}^{N_{\rm\scriptscriptstyle L}-1}\|K_i I_{si} F_{\rm\scriptscriptstyle L}^{N_{\rm\scriptscriptstyle L}-r-1}(A_{\rm\scriptscriptstyle L}-A_{\rm\scriptscriptstyle L}^{\rm\scriptscriptstyle D})\|\sqrt{\sum_{j=1}^{M}\rho^2_{\delta \hat{x}_j}(r-1)}\\
			&\leq\sum_{r=2}^{N_{\rm\scriptscriptstyle L}-1}\|K_i I_{si} F_{\rm\scriptscriptstyle L}^{N_{\rm\scriptscriptstyle L}-r-1}(A_{\rm\scriptscriptstyle L}-A_{\rm\scriptscriptstyle L}^{\rm\scriptscriptstyle D})\|\sum_{j=1}^{M}\rho_{\delta \hat{x}_j}(r-1) \\
			& =\sum_{r=2}^{N_{\rm\scriptscriptstyle L}-1}\|K_i I_{si} F_{\rm\scriptscriptstyle L}^{N_{\rm\scriptscriptstyle L}-r-1}(A_{\rm\scriptscriptstyle L}-A_{\rm\scriptscriptstyle L}^{\rm\scriptscriptstyle D})\|\sum_{j=1}^{M}\sum_{k=1}^{r-1}\|(A_{\rm\scriptscriptstyle L}^{jj})^{r-1-k}B_{\rm\scriptscriptstyle L}^{jj}\|\rho_{\delta \hat{u}_j}\\
			&=\sum_{j=1}^{M}\lambda_{ij}\rho_{\delta \hat{u}_j}
		\end{array}
	\end{equation}
	where $\lambda_{ij}=\sum_{r=2}^{N_{\rm\scriptscriptstyle L}-1}\|K_i I_{si} F_{\rm\scriptscriptstyle L}^{N_{\rm\scriptscriptstyle L}-r-1}(A_{\rm\scriptscriptstyle L}-A_{\rm\scriptscriptstyle L}^{\rm\scriptscriptstyle D})\|\sum_{k=1}^{r-1}\|(A_{\rm\scriptscriptstyle L}^{jj})^{r-1-k}B_{\rm\scriptscriptstyle L}^{jj}\|$. This implies that
	$\Delta \bar{\mathcal{U}}_i\subseteq\mathcal{B}_{\rho_{\delta \hat{u}_i}+\sum_{j=1}^{M}\lambda_{ij}\rho_{\delta \hat{u}_j}}(0)$. Therefore, to verify~\eqref{eq:bound     u_i} it is sufficient to enforce the constraint
	\begin{equation}\label{eq:c de u hat}
		({\Lambda}+I){\overrightarrow{\rho}_{\delta\hat{u}}}+{\overrightarrow{\rho}_{\bar{u}}}\leq {\overrightarrow{\rho}_{{u}}}
	\end{equation}
	where ${\Lambda}$ is the $M\times M$ matrix whose entries are $\lambda_{ij}$, $i,j=1,\dots,M$, while $\overrightarrow{\rho}_{{u}}=(\rho_{u_1},\dots,\rho_{u_{\rm\scriptscriptstyle M}})$, where $\mathcal{B}_{\rho_i}(0)\subseteq \mathcal{U}_{{\rm\scriptscriptstyle S},i}$ for all $i=1,\dots,M$.
	Eventually, a suitable choice of $\overrightarrow{\rho}_{\delta\hat{u}}$ and $\overrightarrow{\rho}_{\bar{u}}$ is obtained as the solution to the following problem:
	\begin{equation}
		\begin{array}{cll}
			\mbox{max} &  & J_{\rho}\\
			\overrightarrow{\rho}_{\delta\hat{u}},\overrightarrow{\rho}_{\bar{u}}\\
			\mbox{subject to}
			&  & \rm{constraint}\,\, \eqref{eq:rho_delta_u_2} \,\,\rm{and}\,\, \eqref{eq:c de u hat}
		\end{array}\label{Eqn:linearoptmiz_u}
	\end{equation}
	where $J_{\rho}$ is a suitable (linear or quadratic, if possible) cost function that allows to maximize the size of the constraint set.
	\subsection{Proof of Theorem \ref{theorem:overall_feasibility}}
	\label{app:proof}
	The proof of Theorem \ref{theorem:overall_feasibility} lies on the intermediary results stated below.
	\begin{proposition}\label{prop:feasibility1}\hfill{}\\
		A) Under Assumption~\ref{Assump:term_constr} and if $\bar{x}^{[N_{\rm\scriptscriptstyle L}]}(k)=\beta x(kN_{\rm\scriptscriptstyle L})$,
		then for any initial condition $\hat{x}(kN_{\rm\scriptscriptstyle L})=x(kN_{\rm\scriptscriptstyle L})$ such that, for all $i=1,\dots,M$
		\begin{equation}
			\|x(kN_{\rm\scriptscriptstyle L})-x_{\rm\scriptscriptstyle S}\|\leq \lambda_{i}(N_{\rm\scriptscriptstyle L})\label{Eqn:condition1}
		\end{equation}
		and for any $\bar{u}^{{[N_{\rm\scriptscriptstyle L}]}}\in\bar{u}_{\rm\scriptscriptstyle S}\oplus{\bar{\mathcal{U}}_{\rm\scriptscriptstyle S}}$
		there exists a feasible sequence\break $\overrightarrow{\delta \hat{u}^{[\zeta_i]}_i} {(kN_i:(k+1)N_i-1|kN_{\rm\scriptscriptstyle L})}\in{\Delta\hat{\mathcal{U}}_{i}}^{N_i}$
		such that the terminal constraint~\eqref{eq:terminaln} is satisfied.\\
		B) if $x(kN_{\rm\scriptscriptstyle L})$ satisfies condition~\eqref{Eqn:condition1}, $\|A_{\rm\scriptscriptstyle L}^{N_{\rm\scriptscriptstyle L}}\|<1$,
		and, for all $i=1,\dots,M$, \eqref{Eqn:chidef} is verified, then recursive feasibility of the terminal constraint~\eqref{eq:terminaln} is guaranteed.
	\end{proposition}

	\noindent
	\emph{Proof of Proposition \ref{prop:feasibility1}}\\\\
	A) Note that,
	since $\delta\hat{x}_{i}^{[\zeta_i]}(kN_i)=\delta\hat{x}_{i}(kN_{\rm\scriptscriptstyle L})=0$,
	\begin{equation}
	\begin{array}{rl}
		&\beta_{i}\delta\hat{x}_{i}((k+1)N_{\rm\scriptscriptstyle L})=
		\beta_{i}\delta\hat{x}_{i}^{[\zeta_i]}((k+1)N_i)=\\
		&={\mathcal{H}}_{i}(N_i)\overrightarrow{\delta \hat{u}^{[\zeta_i]}_i} {(kN_i:(k+1)N_i-1|kN_{\rm\scriptscriptstyle L})}\label{Eqn:rp1}
		\end{array}
	\end{equation}
	Moreover, in view of~\eqref{Eqn:C HLS model}
	\begin{subequations}
		\begin{align}
			\bar{x}^{[N_{\rm\scriptscriptstyle L}]}(k+1)&=A_{\rm\scriptscriptstyle H}^{N_{\rm\scriptscriptstyle L}}\beta x(kN_{\rm\scriptscriptstyle L})+\sum_{j=1}^{N_{\rm\scriptscriptstyle L}}A_{\rm\scriptscriptstyle H}^{N_{\rm\scriptscriptstyle L}-j}B_{\rm\scriptscriptstyle H}\bar{u}^{[N_{\rm\scriptscriptstyle L}]}(k)\label{Eqn:rp2}\\
			\bar{x}_{\rm\scriptscriptstyle S}&=A_{\rm\scriptscriptstyle H}^{N_{\rm\scriptscriptstyle L}}\beta {x}_{\rm\scriptscriptstyle S}+\sum_{j=1}^{N_{\rm\scriptscriptstyle L}}A_{\rm\scriptscriptstyle H}^{N_{\rm\scriptscriptstyle L}-j}B_{\rm\scriptscriptstyle H}\bar{u}_{\rm\scriptscriptstyle S}\label{Eqn:rp2b}
		\end{align}
	\end{subequations}
	Analogously, from~\eqref{Eqn:LL sdyn}
	written in collective form
	\begin{subequations}
		\begin{align}
			\beta\hat{x}(kN_{\rm\scriptscriptstyle L}+N_{\rm\scriptscriptstyle L})&=\beta A_{\rm\scriptscriptstyle L}^{N_{\rm\scriptscriptstyle L}}x(kN_{\rm\scriptscriptstyle L})+\beta\sum_{j=1}^{N_{\rm\scriptscriptstyle L}}A_{\rm\scriptscriptstyle L}^{N_{\rm\scriptscriptstyle L}-j}B_{\rm\scriptscriptstyle L}\bar{u}^{[N_{\rm\scriptscriptstyle L}]}(k)\label{Eqn:rp3}\\
			\beta {x}_{\rm\scriptscriptstyle S}&=\beta A_{\rm\scriptscriptstyle L}^{N_{\rm\scriptscriptstyle L}} {x}_{\rm\scriptscriptstyle S}+\beta\sum_{j=1}^{N_{\rm\scriptscriptstyle L}}A_{\rm\scriptscriptstyle L}^{N_{\rm\scriptscriptstyle L}-j}B_{\rm\scriptscriptstyle L}\bar{u}_{\rm\scriptscriptstyle S}\label{Eqn:rp2c}
		\end{align}
	\end{subequations}
	Therefore, in view of \eqref{Eqn:rp1}, \eqref{Eqn:rp2}, \eqref{Eqn:rp3}, and the definition
	${\mathcal{A}}(N_{\rm\scriptscriptstyle L})$, ${\mathcal{B}}(N_{\rm\scriptscriptstyle L})$, $I_{si}$, the constraint~\eqref{eq:terminaln}
	can be written as
	\begin{equation}
		\begin{array}{rl}
			&{\mathcal{H}}_{i}(N_i)\overrightarrow{\delta \hat{u}^{[\zeta_i]}_i} {(kN_i:(k+1)N_i-1|kN_{\rm\scriptscriptstyle L})}		
			=\\
			&=I_{si}[{\mathcal{A}}(N_{\rm\scriptscriptstyle L})(x(kN_{\rm\scriptscriptstyle L})-x_{\rm\scriptscriptstyle S})
			+\mathcal{B}(N_{\rm\scriptscriptstyle L})(\bar{u}^{[N_{\rm\scriptscriptstyle L}]}(k)-\bar{u}_{\rm\scriptscriptstyle S})]\end{array}\label{Eqn:p4}
	\end{equation}
	From this expression, the definitions of $\underline{\sigma}_{\mathcal{H}_{i}(N_i)}$,
	$\rho_{\bar{u}}$, $\rho_{\delta\hat{u}_{i}}$, and in view of \eqref{Eqn:Gkappa1},
	it can be concluded that a feasible sequence $\overrightarrow{\delta \hat{u}^{[\zeta_i]}_i} {(kN_i:(k+1)N_i-1|kN_{\rm\scriptscriptstyle L})}$
	can be computed provided that
	\begin{equation}
		\sqrt{N_i}\underline{\sigma}_{\mathcal{H}_{i}(N_i)}\rho_{\delta\hat{u}_{i}}\geq\|\mathcal{A}(N_{\rm\scriptscriptstyle L})\|\|x(kN_{\rm\scriptscriptstyle L})-x_{\rm\scriptscriptstyle S}\|+\kappa(N_{\rm\scriptscriptstyle L})\rho_{\bar{u}}\label{Eqn:p5}
	\end{equation}
	from which the result follows.\\\\
	B) From \eqref{Eqn:C full model} it holds that
	\begin{equation}
	\begin{array}{rl}
		&x((k+1)N_{\rm\scriptscriptstyle L})-x_{\rm\scriptscriptstyle S}=A_{\rm\scriptscriptstyle L}^{N_{\rm\scriptscriptstyle L}}\left(x(kN_{\rm\scriptscriptstyle L})-x_{\rm\scriptscriptstyle S}\right)\\
		&+{\mathcal{R}(N_{\rm\scriptscriptstyle L})}
		\left(
		\overrightarrow{u}(kN_{\rm\scriptscriptstyle L}:kN_{\rm\scriptscriptstyle L}+N_{\rm\scriptscriptstyle L}-1|kN_{\rm\scriptscriptstyle L})-\mathds{1}_{N_{\rm\scriptscriptstyle L}\times 1}\otimes\bar{u}_{\rm\scriptscriptstyle S}
		\right)
		\end{array}\label{Eqn:p6}
	\end{equation}
	where $\otimes$ is the Kronecker product.
	Therefore
	\begin{equation}
		\|x((k+1)N_{\rm\scriptscriptstyle L})-x_{\rm\scriptscriptstyle S}\|\leq\|A_{\rm\scriptscriptstyle L}^{N_{\rm\scriptscriptstyle L}}\|\|x(kN_{\rm\scriptscriptstyle L})-x_{\rm\scriptscriptstyle S}\|+\sqrt{N_{\rm\scriptscriptstyle L}}\|{\mathcal{R}(N_{\rm\scriptscriptstyle L})}\|\varrho_{u}\label{Eqn:p7}
	\end{equation}
	and, in view of~\eqref{Eqn:condition1}
	\begin{equation}
		\begin{array}{rl}
			&\|x((k+1)N_{\rm\scriptscriptstyle L})-x_{\rm\scriptscriptstyle S}\|\leq\\
			&\leq\|A_{\rm\scriptscriptstyle L}^{N_{\rm\scriptscriptstyle L}}\|{\frac{(\sqrt{N_i}\underline{\sigma}_{\mathcal{H}_{i}(N_i)}\rho_{\delta \hat{u}_{i}}-\kappa(N_{\rm\scriptscriptstyle L})\rho_{\bar{u}})}{\|{\mathcal{A}}(N_{\rm\scriptscriptstyle L})\|}}+\sqrt{N_{\rm\scriptscriptstyle L}}\|{\mathcal{R}(N_{\rm\scriptscriptstyle L})}\|\varrho_{u}\label{Eqn:p8}
	\end{array}\end{equation}
	for all $i=1,\dots,M$. From this expression and Assumption~\ref{Assump:term_constr}.(4)
	\begin{equation}
		\|x((k+1)N_{\rm\scriptscriptstyle L})-x_{\rm\scriptscriptstyle S}\|\leq{\frac{(\sqrt{N_i}\underline{\sigma}_{\mathcal{H}_{i}(N_i)}\rho_{\delta \hat{u}_{i}}-\kappa(N_{\rm\scriptscriptstyle L})\rho_{\bar{u}})}{\|{\mathcal{A}}(N_{\rm\scriptscriptstyle L})\|}}=\lambda_i(N_L)\label{Eqn:p9}
	\end{equation}
	for all $i=1,2,\dots,M$ and the result follows.\hfill$\square$\\\\
	
	\medskip
	\begin{proposition}\label{prop:prop2} If Problem \eqref{Eqn:LLoptimiz_1} is feasible at time $h=kN_{\rm\scriptscriptstyle L}$, then\\\\
		\noindent
		\begin{align}\|\bar{w}(k)\|\leq \rho_w\label{Eqn:bound_w_2}\\
                    \delta u_i(kN_{\rm\scriptscriptstyle L}+j)\in\Delta\mathcal{U}_i(j)\label{Eqn:bound_deltaui}\end{align}
		Also it holds that
		\begin{equation}\Delta\bar{\mathcal{U}}_i\supseteq \Delta\mathcal{U}_i(j)\label{Eqn:bound_deltaui_monotonicity}\end{equation}
		for all $j=0,\dots,N_{\rm\scriptscriptstyle L}-1$.\hfill$\square$
	\end{proposition}
	
	\noindent
	\emph{Proof of Proposition \ref{prop:prop2}}\\\\
	\noindent
	Defining the collective vectors $\hat{x}=(\hat{x}_1,\dots,\hat{x}_{\rm\scriptscriptstyle M})$, $\delta x=(\delta x_1,\dots,\delta x_{\rm\scriptscriptstyle M})$,\break $\delta \hat{x}=(\delta \hat{x}_1,\dots,\delta \hat{x}_{\rm\scriptscriptstyle M})$, and $\varepsilon(kN_{\rm\scriptscriptstyle L}+j|kN_{\rm\scriptscriptstyle L})=\delta x(kN_{\rm\scriptscriptstyle L}+j|kN_{\rm\scriptscriptstyle L})-\delta \hat{x}(kN_{\rm\scriptscriptstyle L}+j|kN_{\rm\scriptscriptstyle L})$, we have that $\bar{w}(k)=\beta x(kN_{\rm\scriptscriptstyle L}+N_{\rm\scriptscriptstyle L})- \bar{x}^{[N_{\rm\scriptscriptstyle L}]}(k+1|k)=\beta \hat{x}(kN_{\rm\scriptscriptstyle L}+N_{\rm\scriptscriptstyle L})+\beta \delta x(kN_{\rm\scriptscriptstyle L}+N_{\rm\scriptscriptstyle L})- \bar{x}^{[N_{\rm\scriptscriptstyle L}]}(k+1|k)=(\beta \hat{x}(kN_{\rm\scriptscriptstyle L}+N_{\rm\scriptscriptstyle L})+\beta \delta \hat{x}(kN_{\rm\scriptscriptstyle L}+N_{\rm\scriptscriptstyle L})- \bar{x}^{[N_{\rm\scriptscriptstyle L}]}(k+1|k))+\beta\varepsilon(kN_{\rm\scriptscriptstyle L}+j|kN_{\rm\scriptscriptstyle L}) = \beta\varepsilon(kN_{\rm\scriptscriptstyle L}+j|kN_{\rm\scriptscriptstyle L})$. The latter equality holds in view of the fact that Problem \eqref{Eqn:LLoptimiz_1} is feasible, and therefore equality \eqref{eq:terminaln} is verified.
	From \eqref{Eqn:LL ddyn}, \eqref{Eqn:LL dndyn}, \eqref{Eqn:LL dndyn_sf_deltau}, we collectively have that
	\begin{equation}\begin{array}{c}\varepsilon(kN_{\rm\scriptscriptstyle L}+j+1|kN_{\rm\scriptscriptstyle L})=F_{\rm\scriptscriptstyle L} \varepsilon(kN_{\rm\scriptscriptstyle L}+j|kN_{\rm\scriptscriptstyle L})\\
			+ (A_{\rm\scriptscriptstyle L}-A_{\rm\scriptscriptstyle L}^{\rm\scriptscriptstyle D})\delta\hat{x}(kN_{\rm\scriptscriptstyle L}+j|kN_{\rm\scriptscriptstyle L})\end{array}\label{Eqn:epsilon_evolution}\end{equation}
	In view of the fact that $\varepsilon(kN_{\rm\scriptscriptstyle L}|kN_{\rm\scriptscriptstyle L})=\delta \hat{x}(kN_{\rm\scriptscriptstyle L}|kN_{\rm\scriptscriptstyle L})=0$, then
	$\bar{w}(k)=\break\beta \sum_{j=2}^{N_{\rm\scriptscriptstyle L}} F_{\rm\scriptscriptstyle L}^{N_{\rm\scriptscriptstyle L}-j}(A_{\rm\scriptscriptstyle L}-A_{\rm\scriptscriptstyle L}^{\rm\scriptscriptstyle D})\delta\hat{x}(kN_{\rm\scriptscriptstyle L}+j-1|kN_{\rm\scriptscriptstyle L})$. From this it follows that
	\begin{equation}\|\bar{w}(k)\|\leq \sum_{j=2}^{N_{\rm\scriptscriptstyle L}} \|\beta F_{\rm\scriptscriptstyle L}^{N_{\rm\scriptscriptstyle L}-j}(A_{\rm\scriptscriptstyle L}-A_{\rm\scriptscriptstyle L}^{\rm\scriptscriptstyle D})\| \|\delta\hat{x}(kN_{\rm\scriptscriptstyle L}+j-1|kN_{\rm\scriptscriptstyle L})\|\label{Eqn:bound_w_1}\end{equation}
	Since $\delta \hat{u}_{i}$ are bounded for all $i=1,\dots,M$, i.e., scalar $\rho_{\delta \hat{u}_i}$ are defined such that $\delta\hat{u}_{i}\in\mathcal{B}_{\rho_{\delta \hat{u}_i}}(0)$. In view of this, we compute that
	$\|\delta\hat{x}(kN_{\rm\scriptscriptstyle L}+j|kN_{\rm\scriptscriptstyle L})\|\leq \rho_{\delta\hat{x}}(j)$, where $\rho_{\delta\hat{x}_i}(j)$ is defined in \eqref{Eqn:bound_deltaxhat}. Therefore, $\delta\hat{x}(kN_{\rm\scriptscriptstyle L}+j|kN_{\rm\scriptscriptstyle L})$ are bounded, for all $j=1,\dots,N_{\rm\scriptscriptstyle L}-1$ and more specifically we get that $\|\delta\hat{x}(kN_{\rm\scriptscriptstyle L}+j|kN_{\rm\scriptscriptstyle L})\|\leq\break \sqrt{\sum_{i=1}^M\rho^2_{\delta\hat{x}_i}(j)}$.
	Therefore one has \eqref{Eqn:bound_w_2} for all $k\geq 0$.\\\\
	From \eqref{Eqn:epsilon_evolution} we have that $\varepsilon(kN_{\rm\scriptscriptstyle L}+j|kN_{\rm\scriptscriptstyle L})=\sum_{r=2}^j F_{\rm\scriptscriptstyle L}^{j-r}(A_{\rm\scriptscriptstyle L}-A_{\rm\scriptscriptstyle L}^{\rm\scriptscriptstyle D})\delta\hat{x}(kN_{\rm\scriptscriptstyle L}+r-1|kN_{\rm\scriptscriptstyle L})$ and therefore
	$\delta u_i(kN_{\rm\scriptscriptstyle L}+j)-\delta \hat{u}_i(kN_{\rm\scriptscriptstyle L}+j|kN_{\rm\scriptscriptstyle L})=K_i \varepsilon_i(kN_{\rm\scriptscriptstyle L}+j|kN_{\rm\scriptscriptstyle L})=\break K_i I_{si}\sum_{r=2}^j F_{\rm\scriptscriptstyle L}^{j-r}(A_{\rm\scriptscriptstyle L}-A_{\rm\scriptscriptstyle L}^{\rm\scriptscriptstyle D})\delta\hat{x}(kN_{\rm\scriptscriptstyle L}+r-1|kN_{\rm\scriptscriptstyle L})$. From this it follows that
	$\delta u_i(kN_{\rm\scriptscriptstyle L}+j)\in\delta \hat{u}_i(kN_{\rm\scriptscriptstyle L}+j|kN_{\rm\scriptscriptstyle L})\oplus \mathcal{B}_{\rho_{\Delta{u}_i}(j)}(0)$ and therefore $\delta u_i(kN_{\rm\scriptscriptstyle L}+j)\in\Delta\mathcal{U}_i(j)$. In view of the monotonicity property $\rho_{\Delta{u}_i}(j+1)\geq \rho_{\Delta{u}_i}(j)$ for all $j$, it holds that\break $\mathcal{B}_{\rho_{\Delta{u}_i}(j+1)}(0)\supseteq \mathcal{B}_{\rho_{\Delta{u}_i}(j)}(0)$, which implies \eqref{Eqn:bound_deltaui_monotonicity}.\hfill$\square$\\

	\noindent
	\emph{Proof of Theorem \ref{theorem:overall_feasibility}}\\\\
	(i) If $\|x(0)-x_{\rm\scriptscriptstyle S}\|\leq \lambda_{i}(N_{\rm\scriptscriptstyle L})$ and recalling that Assumption~\ref{Assump:term_constr} holds, from Proposition~\ref{prop:feasibility1}, recursive feasibility of the optimization problems~\eqref{Eqn:LLoptimiz_1} is guaranteed, i.e.,  that there exists, for all $k\geq 0$, a feasible sequence $\overrightarrow{\delta \hat{u}^{[\zeta_i]}_i} {(kN_i:(k+1)N_i-1|kN_{\rm\scriptscriptstyle L})}\in{\Delta\hat{\mathcal{ U}}_{i}}^{N_i}$
	such that the terminal constraint~\eqref{eq:terminaln} is satisfied.\\
	Also, from Proposition \ref{prop:prop2}, it follows that $\bar{w}(k)\in\mathcal{W}$ for all $k\geq 0$, which allows to apply the recursive feasibility arguments of \cite{Mayne2005219}, proving that also \eqref{Eqn:HLoptimiz_1} enjoys recursive feasibility properties.\\
	(ii) It is now possible to conclude that, in view of the feasibility of \eqref{Eqn:HLoptimiz_1}, $\bar{u}^{[N_{\rm\scriptscriptstyle L}]}(k)\in\bar{u}_{\rm\scriptscriptstyle S}\oplus\bar{\mathcal{U}}_{\rm\scriptscriptstyle S}$; also, from Proposition~\ref{prop:prop2} it follows that $\delta u_{i}(kN_{\rm\scriptscriptstyle L}+j)\in \Delta\bar{\mathcal{U}}_i$ for all $k\geq 0$, $j=0,\dots, N_{\rm\scriptscriptstyle L}-1$, and $i=1,\dots,M$. From this, under \eqref{eq:bound u_def}, the inclusion~\eqref{eq:bound u} can also be proved.\\
	(iii) We apply the results in \cite{Mayne2005219}, which guarantee robust convergence properties. In other words, it holds that $\bar{x}^{[N_{\rm\scriptscriptstyle L}],o}(k)\rightarrow \bar{x}_{\rm\scriptscriptstyle S}$ as $k\rightarrow +\infty$, and that $\bar{x}^{[N_{\rm\scriptscriptstyle L}]}(k)$ is asymptotically driven to lie in the robust positively invariant set $\bar{x}_{\rm\scriptscriptstyle S}\oplus\mathcal{Z}$.\\
	(iv) To show robust convergence of the global system state, from \eqref{Eqn:C full model} we obtain that
	\begin{align}x((k+1)N_{\rm\scriptscriptstyle L})=A_{\rm\scriptscriptstyle L}^{N_{\rm\scriptscriptstyle L}}x(kN_{\rm\scriptscriptstyle L})+B_{\rm\scriptscriptstyle L}^{[N_{\rm\scriptscriptstyle L}]}\bar{u}^{[N_{\rm\scriptscriptstyle L}]}(k)\nonumber\\
		+\sum_{h=0}^{N_{\rm\scriptscriptstyle L}-1}A_{\rm\scriptscriptstyle L}^hB_{\rm\scriptscriptstyle L}\delta u((k+1)N_{\rm\scriptscriptstyle L}-h-1)\label{eq:evol_x_contr01}\end{align}
	Denoting $\mathcal{B}_{{\rm\scriptscriptstyle L},N_{\rm\scriptscriptstyle L}}^{\rm\scriptscriptstyle C}=\begin{bmatrix}A_{\rm\scriptscriptstyle L}^{N_{\rm\scriptscriptstyle L}-1}B_{\rm\scriptscriptstyle L}&\dots &B_{\rm\scriptscriptstyle L}\end{bmatrix}$, we obtain that
	$$\sum_{h=0}^{N_{\rm\scriptscriptstyle L}-1}A_{\rm\scriptscriptstyle L}^hB_{\rm\scriptscriptstyle L}\delta u((k+1)N_{\rm\scriptscriptstyle L}-h-1)=\mathcal{B}_{{\rm\scriptscriptstyle L},N_{\rm\scriptscriptstyle L}}^{\rm\scriptscriptstyle C}\overrightarrow{\delta {u}} {(kN_{\rm\scriptscriptstyle L}:kN_{\rm\scriptscriptstyle L}+N_{\rm\scriptscriptstyle L}-1)}$$
	Also, recall that
	$\overrightarrow{\delta {u}} {(kN_{\rm\scriptscriptstyle L}:kN_{\rm\scriptscriptstyle L}+N_{\rm\scriptscriptstyle L}-1)}=\overrightarrow{\delta \hat{u}} (kN_{\rm\scriptscriptstyle L}:kN_{\rm\scriptscriptstyle L}+N_{\rm\scriptscriptstyle L}-1)+\break{\rm{diag}}(K,\dots,K)\overrightarrow{\varepsilon} {(kN_{\rm\scriptscriptstyle L}:kN_{\rm\scriptscriptstyle L}+N_{\rm\scriptscriptstyle L}-1)}$ and that, by defining
	$$\mathcal{F}_{N_{\rm\scriptscriptstyle L}}=\begin{bmatrix}0&0&\cdots&0&0&0\\
	I&0&\cdots&0&0&0\\
	F_{\rm\scriptscriptstyle L}&I&\cdots&0&0&0\\
	\vdots&\vdots&\ddots&\vdots&\vdots&\vdots\\
	F_{\rm\scriptscriptstyle L}^{N_{\rm\scriptscriptstyle L}-2}&F_{\rm\scriptscriptstyle L}^{N_{\rm\scriptscriptstyle L}-3}&\cdots&I&0&0\end{bmatrix},$$
	then 
	$$\begin{array}{rl}
	&\overrightarrow{\varepsilon} {(kN_{\rm\scriptscriptstyle L}:kN_{\rm\scriptscriptstyle L}+N_{\rm\scriptscriptstyle L}-1)}
	=\\
	&=\mathcal{F}_{N_{\rm\scriptscriptstyle L}}{\rm{diag}}(A_{\rm\scriptscriptstyle L}^{\rm\scriptscriptstyle C},\cdots,A_{\rm\scriptscriptstyle L}^{\rm\scriptscriptstyle C})\mathcal{B}_{\rm\scriptscriptstyle L}\overrightarrow{\delta \hat{u}} {(kN_{\rm\scriptscriptstyle L}:kN_{\rm\scriptscriptstyle L}+N_{\rm\scriptscriptstyle L}-1)}
	\end{array}$$
	where $A_{\rm\scriptscriptstyle L}^{\rm\scriptscriptstyle C}=A_{\rm\scriptscriptstyle L}-A_{\rm\scriptscriptstyle L}^{\rm\scriptscriptstyle D}$ and
	$$\mathcal{B}_{\rm\scriptscriptstyle L}=\begin{bmatrix}
	0&\cdots&0\\
	\vdots&\ddots&\vdots\\
	(A_{\rm\scriptscriptstyle L}^{\rm\scriptscriptstyle D})^{N_{\rm\scriptscriptstyle L}-1}B_{\rm\scriptscriptstyle L}&\cdots&B_{\rm\scriptscriptstyle L}\end{bmatrix}$$
	Recalling that $\bar{x}^{[N_{\rm\scriptscriptstyle L}]}(k)=\beta x(kN_{\rm\scriptscriptstyle L})$ and that
	\begin{align}\bar{u}^{[N_{\rm\scriptscriptstyle L}]}(k)=\bar{u}^{[N_{\rm\scriptscriptstyle L}],o}(k)+\bar{K}_{\rm\scriptscriptstyle H} (\bar{x}^{[N_{\rm\scriptscriptstyle L}]}(k)-\bar{x}^{[N_{\rm\scriptscriptstyle L}],o}(k))\label{eq:u^NL_TOT01}\end{align}
	we can rewrite \eqref{eq:evol_x_contr01} as
	\begin{align}&x((k+1)N_{\rm\scriptscriptstyle L})=(A_{\rm\scriptscriptstyle L}^{N_{\rm\scriptscriptstyle L}}+B_{\rm\scriptscriptstyle L}^{[N_{\rm\scriptscriptstyle L}]}\bar{K}_{\rm\scriptscriptstyle H}\beta)x(kN_{\rm\scriptscriptstyle L})\nonumber\\
		&+B_{\rm\scriptscriptstyle L}^{[N_{\rm\scriptscriptstyle L}]}(\bar{u}^{[N_{\rm\scriptscriptstyle L}],o}(k)-\bar{K}_{\rm\scriptscriptstyle H}\bar{x}^{[N_{\rm\scriptscriptstyle L}],o}(k))+\mathcal{B}_{{\rm\scriptscriptstyle L},N_{\rm\scriptscriptstyle L}}^{\rm\scriptscriptstyle C}\overrightarrow{\delta {u}} {(kN_{\rm\scriptscriptstyle L}:kN_{\rm\scriptscriptstyle L}+N_{\rm\scriptscriptstyle L}-1)}\label{eq:evol_x_contr01bis}\end{align}
	Recall that $\bar{x}^{[N_{\rm\scriptscriptstyle L}],o}(k)\rightarrow\bar{x}_{\rm\scriptscriptstyle S}$, $\bar{u}^{[N_{\rm\scriptscriptstyle L}],o}(k)\rightarrow \bar{u}_{\rm\scriptscriptstyle S}$ as $k\rightarrow +\infty$. Also, we compute that
	\begin{equation}\begin{array}{c}\overrightarrow{\delta {u}} {(kN_{\rm\scriptscriptstyle L}:kN_{\rm\scriptscriptstyle L}+N_{\rm\scriptscriptstyle L}-1)} =(I+{\rm{diag}}(K_i,\dots,K_i)\mathcal{F}_{N_{\rm\scriptscriptstyle L}}\\
			\cdot{\rm{diag}}(A_{\rm\scriptscriptstyle L}^{\rm\scriptscriptstyle C},\dots,A_{\rm\scriptscriptstyle L}^{\rm\scriptscriptstyle C})\mathcal{B}_{\rm\scriptscriptstyle L})\overrightarrow{\delta \hat{u}} {(kN_{\rm\scriptscriptstyle L}:kN_{\rm\scriptscriptstyle L}+N_{\rm\scriptscriptstyle L}-1)}\end{array}\label{eq:deltau_tot_new}\end{equation}
	Based on this, we define $\kappa_{\delta u}=\|\mathcal{B}_{{\rm\scriptscriptstyle L},N_{\rm\scriptscriptstyle L}}^{\rm\scriptscriptstyle C}(I+{\rm{diag}}(K_i,\dots,K_i)\\ \mathcal{F}_{N_{\rm\scriptscriptstyle L}}{\rm{diag}}(A_{\rm\scriptscriptstyle L}^{\rm\scriptscriptstyle C},\dots,A_{\rm\scriptscriptstyle L}^{\rm\scriptscriptstyle C})\mathcal{B}_{\rm\scriptscriptstyle L})\|$ and we write \eqref{eq:evol_x_contr01bis} as
	\begin{align}&x((k+1)N_{\rm\scriptscriptstyle L})=(A_{\rm\scriptscriptstyle L}^{N_{\rm\scriptscriptstyle L}}+B_{\rm\scriptscriptstyle L}^{[N_{\rm\scriptscriptstyle L}]}\bar{K}_{\rm\scriptscriptstyle H}\beta)x(kN_{\rm\scriptscriptstyle L})\label{eq:evol_x_contr01tris}\\
		&+
		B_{\rm\scriptscriptstyle L}^{[N_{\rm\scriptscriptstyle L}]}(\bar{u}^{[N_{\rm\scriptscriptstyle L}],o}(k)-\bar{K}_{\rm\scriptscriptstyle H}\bar{x}^{[N_{\rm\scriptscriptstyle L}],o}(k))+w_{\rm\scriptscriptstyle L}(k)\nonumber\end{align}
	where $\|w_{\rm\scriptscriptstyle L}(k)\|\leq \kappa_{\delta u} \sqrt{N_{\rm\scriptscriptstyle L}} \max_{h\in\{kN_{\rm\scriptscriptstyle L},\dots,(k+1)N_{\rm\scriptscriptstyle L}-1\}}\|\delta \hat{u}(h)\|\leq \kappa_{\delta u} \sqrt{N_{\rm\scriptscriptstyle L}}\sqrt{\sum_{i=1}^M\rho^2_{\delta \hat{u}_i}}$.
	Since $x_{\rm\scriptscriptstyle S}=A_{\rm\scriptscriptstyle L}^{N_{\rm\scriptscriptstyle L}}x_{\rm\scriptscriptstyle S}+B_{\rm\scriptscriptstyle L}^{[N_{\rm\scriptscriptstyle L}]}\bar{u}_{\rm\scriptscriptstyle S}$ and $B_{\rm\scriptscriptstyle L}^{[N_{\rm\scriptscriptstyle L}]}\bar{K}_{\rm\scriptscriptstyle H}\bar{x}_{\rm\scriptscriptstyle S}=B_{\rm\scriptscriptstyle L}^{[N_{\rm\scriptscriptstyle L}]}\bar{K}_{\rm\scriptscriptstyle H}\beta {x}_{\rm\scriptscriptstyle S}$, we can rewrite~\eqref{eq:evol_x_contr01tris} as
	$$\begin{array}{rl}&x((k+1)N_{\rm\scriptscriptstyle L})-x_{\rm\scriptscriptstyle S}=(A_{\rm\scriptscriptstyle L}^{N_{\rm\scriptscriptstyle L}}+B_{\rm\scriptscriptstyle L}^{[N_{\rm\scriptscriptstyle L}]}\bar{K}_{\rm\scriptscriptstyle H}\beta)(x(kN_{\rm\scriptscriptstyle L})-x_{\rm\scriptscriptstyle S})+w_{\rm\scriptscriptstyle L}(k)\\
	&+B_{\rm\scriptscriptstyle L}^{[N_{\rm\scriptscriptstyle L}]}\left((\bar{u}^{[N_{\rm\scriptscriptstyle L}],o}(k)-\bar{u}_{\rm\scriptscriptstyle S})-\bar{K}_{\rm\scriptscriptstyle H}(\bar{x}^{[N_{\rm\scriptscriptstyle L}],o}(k)-\bar{x}_{\rm\scriptscriptstyle S})\right)\end{array}$$
	Eventually, since $F_{\rm\scriptscriptstyle L}^{[N_{\rm\scriptscriptstyle L}]}=A_{\rm\scriptscriptstyle L}^{N_{\rm\scriptscriptstyle L}}+B_{\rm\scriptscriptstyle L}^{[N_{\rm\scriptscriptstyle L}]}\bar{K}_{\rm\scriptscriptstyle H}\beta$ is Schur stable, then the asymptotic result follows, where
	$\rho_x=\kappa_{\delta u} \sqrt{N_{\rm\scriptscriptstyle L}} \sqrt{\sum_{i=1}^M\rho^2_{\delta \hat{u}_i}}$.\\
	(v) We can reformulate problem \eqref{Eqn:LLoptimiz_1} as
	\begin{equation}
		\begin{array}{cll}
			\mbox{min} &  & \|\overrightarrow{\delta \hat{u}} {(kN_{\rm\scriptscriptstyle L}:kN_{\rm\scriptscriptstyle L}+N_{\rm\scriptscriptstyle L}-1)}\|^2_{\mathcal{Q}}\\
			\mbox{subject to:} &  & \beta
			\mathcal{R}^{\rm\scriptscriptstyle D}(N_{\rm\scriptscriptstyle L})
			\overrightarrow{\delta \hat{u}} {(kN_{\rm\scriptscriptstyle L}:kN_{\rm\scriptscriptstyle L}+N_{\rm\scriptscriptstyle L}-1)}=\delta\bar{x}_{end}\\
			&  & A_{in}\overrightarrow{\delta \hat{u}} {(kN_{\rm\scriptscriptstyle L}:kN_{\rm\scriptscriptstyle L}+N_{\rm\scriptscriptstyle L}-1)}\leq b_{in},
		\end{array}\label{Eqn:LLoptimiz_2}
	\end{equation}
	where $\mathcal{Q}=$diag$(R,\dots,R)+\mathcal{B}_{\rm\scriptscriptstyle L}^T$diag$(Q,\dots,Q)\mathcal{B}_{\rm\scriptscriptstyle L}$,
	$R=$diag\\$(R_1,\dots,R_{\rm\scriptscriptstyle M})$, $Q=$diag$(Q_1,\dots,Q_{\rm\scriptscriptstyle M})$, $\delta\bar{x}_{end}=\bar{x}^{[N_{\rm\scriptscriptstyle L}]}(k+1|k)-\beta\hat{x}(kN_{\rm\scriptscriptstyle L}+N_{\rm\scriptscriptstyle L})$ and\break $\mathcal{R}^{\rm\scriptscriptstyle D}(N_{\rm\scriptscriptstyle L})=\begin{bmatrix}
	(A_{\rm\scriptscriptstyle L}^{\rm\scriptscriptstyle D})^{N_{\rm\scriptscriptstyle L}-1}B_{\rm\scriptscriptstyle L}&\dots&B_{\rm\scriptscriptstyle L}\end{bmatrix}$. Recalling that $b_{in}>0$ elementwise, in view of continuity arguments, there exists a ball $\mathcal{B}_{\rho_{end}}(0)$ for $\delta\bar{x}_{end}$ such that the solution to problem \eqref{Eqn:LLoptimiz_2} satisfies $A_{in}\overrightarrow{\delta \hat{u}} {(kN_{\rm\scriptscriptstyle L}:kN_{\rm\scriptscriptstyle L}+N_{\rm\scriptscriptstyle L}-1)}\leq b_{in}$. If
	$\delta\bar{x}_{end}\in\mathcal{B}_{\rho_{end}}(0)$, then
	$$\begin{array}{c}\overrightarrow{\delta \hat{u}} {(kN_{\rm\scriptscriptstyle L}:kN_{\rm\scriptscriptstyle L}+N_{\rm\scriptscriptstyle L}-1)}\\=\begin{bmatrix}I&0\end{bmatrix}\begin{bmatrix}2\mathcal{Q}&(\beta \mathcal{R}^{\rm\scriptscriptstyle D}(N_{\rm\scriptscriptstyle L}))^T\\\beta \mathcal{R}^{\rm\scriptscriptstyle D}(N_{\rm\scriptscriptstyle L})&0
	\end{bmatrix}^{-1}\begin{bmatrix}0\\I\end{bmatrix}\delta\bar{x}_{end}\end{array}$$
	Also, it holds that the optimal constrained solution fulfills
	$$\|\overrightarrow{\delta \hat{u}} {(kN_{\rm\scriptscriptstyle L}:kN_{\rm\scriptscriptstyle L}+N_{\rm\scriptscriptstyle L}-1)}\|^2_{\mathcal{Q}}\leq \max_{A_{in}\overrightarrow{\delta \hat{u}} \leq b_{in}}\|\overrightarrow{\delta \hat{u}} \|^2_{\mathcal{Q}}=J_{\rm\scriptscriptstyle L}^{max}$$
	and therefore $\|\overrightarrow{\delta \hat{u}} {(kN_{\rm\scriptscriptstyle L}:kN_{\rm\scriptscriptstyle L}+N_{\rm\scriptscriptstyle L}-1)}\|\leq \sqrt{\frac{J_{\rm\scriptscriptstyle L}^{max}}{\lambda_{min}(\mathcal{Q})}}$; in view of this, for $\|\delta\bar{x}_{end}\|\not\in \mathcal{B}_{\rho_{end}}(0)$, then $\|\overrightarrow{\delta \hat{u}} (kN_{\rm\scriptscriptstyle L}:kN_{\rm\scriptscriptstyle L}+N_{\rm\scriptscriptstyle L}-1)\|\leq \sqrt{\frac{J_{\rm\scriptscriptstyle L}^{max}}{\lambda_{min}(\mathcal{Q})}}\frac{1}{\rho_{end}}\|\delta\bar{x}_{end}\|$.
	Defining now
	$\bar{\kappa}=\max\{\left\|\begin{bmatrix}I&0\end{bmatrix}\begin{bmatrix}2\mathcal{Q}&(\beta \mathcal{R}^{\rm\scriptscriptstyle D}(N_{\rm\scriptscriptstyle L}))^T\\\beta \mathcal{R}^{\rm\scriptscriptstyle D}(N_{\rm\scriptscriptstyle L})&0
	\end{bmatrix}^{-1}\begin{bmatrix}0\\I\end{bmatrix}\right\|,\sqrt{\frac{J_{\rm\scriptscriptstyle L}^{max}}{\lambda_{min}(\mathcal{Q})}}\frac{1}{\rho_{end}}\}$ then we conclude that
	\begin{equation}
		\|\overrightarrow{\delta \hat{u}} {(kN_{\rm\scriptscriptstyle L}:kN_{\rm\scriptscriptstyle L}+N_{\rm\scriptscriptstyle L}-1)}\|\leq \bar{\kappa}\|\delta\bar{x}_{end}\|
		\label{eq:opt_inequality}
	\end{equation}
	%
	Therefore, from \eqref{eq:deltau_tot_new} and \eqref{eq:opt_inequality}, we have that
	\begin{align}\|\overrightarrow{\delta {u}} {(kN_{\rm\scriptscriptstyle L}:kN_{\rm\scriptscriptstyle L}+N_{\rm\scriptscriptstyle L}-1)}\|\leq \kappa_u \|\overrightarrow{\delta \hat{u}} {(kN_{\rm\scriptscriptstyle L}:kN_{\rm\scriptscriptstyle L}+N_{\rm\scriptscriptstyle L}-1)}\|\nonumber\\
		\leq \kappa_u\bar{\kappa}\|{\mathcal{A}}(N_{\rm\scriptscriptstyle L})(x(kN_{\rm\scriptscriptstyle L})-x_{\rm\scriptscriptstyle S})+\mathcal{B}(N_{\rm\scriptscriptstyle L})(\bar{u}^{[N_{\rm\scriptscriptstyle L}]}(k)-\bar{u}_{\rm\scriptscriptstyle S}) \|\label{eq:deltau_vec}\end{align}
	where
	$$\kappa_u=\|I+{\rm{diag}}(K_i,\dots,K_i)\mathcal{F}_{N_{\rm\scriptscriptstyle L}}{\rm{diag}}(A_{\rm\scriptscriptstyle L}^{\rm\scriptscriptstyle C},\dots,A_{\rm\scriptscriptstyle L}^{\rm\scriptscriptstyle C})\mathcal{B}_{\rm\scriptscriptstyle L}\|$$
	In view of this, we can rewrite \eqref{eq:evol_x_contr01bis} as
	\begin{align}x((k+1)N_{\rm\scriptscriptstyle L})-x_{\rm\scriptscriptstyle S}&=
		F_{\rm\scriptscriptstyle L}^{[N_{\rm\scriptscriptstyle L}]}(x(kN_{\rm\scriptscriptstyle L})-x_{\rm\scriptscriptstyle S})+w_{x}(k)+w_{o}(k)\label{eq:evol_x_contr02}\end{align}
	where $\|w_{x}(k)\|\leq \kappa_x(N_{\rm\scriptscriptstyle L}) \|x(kN_{\rm\scriptscriptstyle L})-x_{\rm\scriptscriptstyle S}\|$, $\kappa_x(N_{\rm\scriptscriptstyle L})=\kappa_u\bar{\kappa}\|\mathcal{B}_{{\rm\scriptscriptstyle L},N_{\rm\scriptscriptstyle L}}^{\rm\scriptscriptstyle C}\|\\\|{\mathcal{A}}(N_{\rm\scriptscriptstyle L})+{\mathcal{B}}(N_{\rm\scriptscriptstyle L})\bar{K}_{\rm\scriptscriptstyle H}\beta\|$, $\|w_o(k)\|\leq \kappa_u^o\|\bar{u}^{[N_{\rm\scriptscriptstyle L}],o}(k)-\bar{u}_{\rm\scriptscriptstyle S}\|+\kappa_x^o\|\bar{x}^{[N_{\rm\scriptscriptstyle L}],o}(k)-\bar{x}_{\rm\scriptscriptstyle S}\|$, $\kappa_u^o=\|B_{\rm\scriptscriptstyle L}^{[N_{\rm\scriptscriptstyle L}]}\|+\kappa_u\bar{\kappa}\|\mathcal{B}_{{\rm\scriptscriptstyle L},N_{\rm\scriptscriptstyle L}}^{\rm\scriptscriptstyle C}\|\|{\mathcal{B}}(N_{\rm\scriptscriptstyle L})\|$, and $$\kappa_x^o=\|B_{\rm\scriptscriptstyle L}^{[N_{\rm\scriptscriptstyle L}]}\bar{K}_{\rm\scriptscriptstyle H}\|+\kappa_u\bar{\kappa}\|\mathcal{B}_{{\rm\scriptscriptstyle L},N_{\rm\scriptscriptstyle L}}^{\rm\scriptscriptstyle C}\|\|{\mathcal{B}}(N_{\rm\scriptscriptstyle L})\bar{K}_{\rm\scriptscriptstyle H}\|$$
	To derive a stability condition, we recast~\eqref{eq:evol_x_contr02} as the following redundant dynamic system
	\begin{equation}\begin{array}{lcl}
			\delta x_1^+&=&
			F_{\rm\scriptscriptstyle L}^{[N_{\rm\scriptscriptstyle L}]}\delta x_1+w_{\delta x_2}+w_{o}\\
			\delta x_2^+&=&
			F_{\rm\scriptscriptstyle L}^{[N_{\rm\scriptscriptstyle L}]}\delta x_2+w_{\delta x_1}+w_{o}
		\end{array}\label{eq:duplicated}\end{equation}
	where the initial conditions for $\delta x_1$ and $\delta x_2$ coincide and are equal to $x(0)-x_{\rm\scriptscriptstyle S}$, and where $\|w_{\delta x_i}\|\leq \kappa_x(N_{\rm\scriptscriptstyle L}) \|\delta x_i\|$, $i=1,2$. Recall also that, as already discussed, $w_{o}$ is an asymptotically vanishing input.\\
	The stability of the system~\eqref{eq:duplicated} can be studied using the (ISS) small gain theorem in \cite{Dashkovskiy2007}, according to which the interconnected system above enjoys asymptotic stability properties if the matrix
	$$\Gamma=\begin{bmatrix}0&\sigma(N_{\rm\scriptscriptstyle L})\\\sigma(N_{\rm\scriptscriptstyle L})&0\end{bmatrix}$$
	is Schur stable, where
	\begin{align}\sigma(N_{\rm\scriptscriptstyle L})=\kappa_x(N_{\rm\scriptscriptstyle L})\sum_{k=0}^{+\infty}\|(F_{\rm\scriptscriptstyle L}^{[N_{\rm\scriptscriptstyle L}]})^{k}\|
		\label{eq:small_gain_final}\end{align}
	Also, $\Gamma$ is Schur stable if and only if the inequality \eqref{eq:small_gain_final2} is verified.
	Then, under the latter condition, $x(kN_{\rm\scriptscriptstyle L})\rightarrow x_{\rm\scriptscriptstyle S}$ as $k\rightarrow +\infty$.	 \hfill{}$\square$\\\\
	\bibliographystyle{plain}        
	\bibliography{stochasticMPC}
	%
\end{document}